\documentclass{aa}  

\usepackage{color, colortbl}
\usepackage{graphicx}
\usepackage{tikz}
\usepackage{float}
\usepackage{subcaption}
\usepackage{multicol}
\usepackage{multirow}
\usepackage{longtable}
\usepackage{placeins}
\usepackage{hyperref}
\usepackage{soul}
\usepackage{txfonts}

\def\LP#1{}
\def\AJ#1{}
\def\LP#1{{\color{blue} \textsl{\small $^{[LP]}[$#1$]$}}} 
\def\AJ#1{{\color{red} \textsl{\small $^{[AJ]}[$#1$]$}}} 

\begin{document} 

   \title{Spectroscopic monitoring of long secondary period variables}

   \subtitle{The case of J-type carbon stars\thanks{Based on observations made with the Mercator Telescope, operated on the island of La Palma by the Flemish Community, at the Spanish Observatorio del Roque de los Muchachos of the Instituto de Astrof\'isica de Canarias.}}

\author{L. Planquart
          \inst{1}\and A. Jorissen\inst{2,3}    
          \and H. Van Winckel\inst{4}}

   \institute{Department of Physics and Astronomy,
Chalmers University of Technology,
SE-412 96 Gothenburg, Sweden
\and 
Institut d’Astronomie et d’Astrophysique, Université libre de Bruxelles, CP 226, Boulevard du Triomphe, 1050 Brussels, Belgium
\and
BLU-ULB, Brussels Laboratory of the Universe, https://blu.ulb.be
\and 
Institute of Astronomy, KU Leuven, Celestijnenlaan 200D, 3001 Leuven, Belgium
             }

   \date{Received 15/07/2026 / Accepted 25/09/2026}

 
  \abstract
  {Long-secondary-period (LSP) variables are a subclass of long-period variables whose origin is a matter of debate. J-type carbon stars are a subclass of carbon stars enriched in $\rm ^{13}C$, and the origin of this chemical peculiarity is unclear. }
   {We report on the discovery of a LSP in two J-type carbon stars: VZ~Per and UV~Cam.}
   {A decade-long spectroscopic monitoring was used to obtain a long-term radial-velocity curve for the two stars. From a frequency analysis, we isolated a LSP signal. The retrieved periodicity was then compared with changes in spectral profiles and available light-curve variability to further constrain the origin of the modulation.}
   {For VZ Per, the LSP derived from the radial-velocity curve ($\sim\,$960 days) is correlated with the light-curve dimming event and a variation in the Balmer absorption line profile. These results translate into changes in the gas and dust column densities with the phase of the LSP, with the maximum gas column density occurring ahead of the dust. For UV~Cam, the radial velocity monitoring yields a period of $\sim\,1178$ d with an amplitude of about 1\,km/s. Interpreting the RV modulation as a Keplerian signal for both stars, a low-mass ($<0.1\, \rm M_\odot$) close companion is found. In this scenario, variations in gas and dust column densities with (orbital) phase trace the material in the expanding comet-like tail trailing the substellar companion. }
   {These observations provide time-dependent variations of the gas and dust in LSP variables that are consistent with the current binary scenario. Furthermore, the detection of LSP among J-type carbon stars anchors their current evolutionary status in the dust-producing region of the thermally-pulsing AGB phase.}


   \maketitle

\section{Introduction} \label{sect:intro}

About one-third of long-period variables among red giants and supergiants exhibit a long secondary period (LSP) in their light curve. With periods ranging from 300 to 1500 days, their positions in the Period-Luminosity (PL) diagram define the sequence D, parallel to the pulsating-star sequences (labeled from A to C) and roughly in the continuation of the sequence E, populated by eclipsing binaries, at higher luminosities and longer periods \citep{1996MNRAS.282..958W}.

Over the years, several scenarios have been proposed to explain this distinct variability class: dust ejections, stellar rotation, (non-)radial pulsations, and dust-enshrouded companions \citep{Wood_LSP_spectro_2004ApJ...604..800W}. Complementary observations -- mainly radial-velocity (RV) monitoring \citep{Nicholls_2009MNRAS.399.2063N} and infrared light curves \citep{Soszynski_2021ApJ...911L..22S} -- have placed strong constraints on the physical model required to explain the LSP and challenge the various scenarios that have to deal with the shape of the RV curve and the phase lag between the RV and light curves in different photometric bands (optical and near-infrared). 
The recent Gaia data releases \citep{2023A&A...674A...1G} have reignited the current tension about the LSP origin: \cite{2026PASP..138e4202S} have stated that Gaia DR3 astrometry -- especially the \textsc{ruwe} value -- disfavors the binary origin of LSP, but the statement was later challenged by \cite{2026arXiv260605292I}, who perform an independent reassessment of a LSP sample present in Gaia.

While the origin of LSP remains elusive, two leading hypotheses to explain its variability have attracted attention because they can, at least partially, account for these observational constraints.
On one side, an intrinsic origin caused by a non-radial pulsation, such as an oscillatory convective mode \citep{LSP_dipole_oscillation_2015MNRAS.452.3863S}. 
In this scenario, a change in radius is responsible for the RV variation, while a change in effective temperature ($\sim150$\,K) could account for the color variation in the light curve \citep{LSP_dipole_mode_2020MNRAS.492.1348T}.
Such an intrinsic phenomenon seems favored by the lack of asymmetric signals at long wavelengths in interferometric observations of RT Pav, a close and bright star with a LSP \citep{LSP_XT_Pav_2026A&A...705A.187C}.

On the other hand, a compelling scenario involves an extrinsic origin caused by dust cometary tails following an orbiting low-mass companion \citep{Soszynski_2021ApJ...911L..22S}. 
In this scenario, the color variation would be explained by dust extinction and would provide a natural explanation for infrared secondary eclipses, while orbital motion due to a substellar companion would explain the low-amplitude RV curve variation. 
The non-sinusoidal shape of the RV-curves seems to be biased toward eccentric orbits \citep{Nicholls_2009MNRAS.399.2063N}, a fact that has been seen as a limitation of the binary-induced scenario. However, it can be overcome by introducing either a distortion effect of the orbiting dust cloud in a circular orbit \citep{2014ApJ...788...13S} or a time-in-dust selection effect in an eccentric orbit \citep{Decin_2025A&A...703L..23D}. In the latter, there is a geometric selection effect in favor of low-mass companions in close eccentric orbits where the companion only spends part of its orbit within the giant's dust-formation zone. 
In the case of a supergiant like Betelgeuse, the detection of a binary companion responsible for the LSP through secular radial-velocity (RV) and astrometric variations (\citealt{2024ApJ...977...35G}, \citealt{2025ApJ...978...50M}) provides a strong argument in favor of the binary scenario for LSP extendable to their lower-mass counterparts, the long-period variables. If the above scenario is correct, the presence of a substellar companion would impact the dust and gas distribution around the evolved star and affect the overall dust budget \citep{2025A&A...704A.341D}.

The chemical composition of the wind of a star on the Asymptotic Giant Branch (AGB) is either dominated by oxygen-bearing molecules and silicate-rich dust or carbonaceous molecules and carbon-rich mineralogy. 
Such a dichotomy arises from an evolutionary sequence: as the star evolves through the AGB phase, it transforms an oxygen-rich ($\rm C/O<1$) into a carbon-rich ($\rm C/O>1$) atmosphere by enriching it in $\rm ^{12}C$ and $s$ process elements.
J-type stars constitute a subclass of carbon stars (about 15\% of them) known to be enriched in $\rm ^{13}C$, leading to $\rm ^{12}C/^{13}C<15$ \citep{temperature_1999A&A...345..233O}. Their optical spectra also often exhibit a strong lithium line and lack $s$-process or Tc lines \citep{Abia_Tc_2000ApJ...536..438A}. The $\rm ^{13}C$ overabundance challenges standard AGB models, indicating that J-type stars may follow a different evolutionary sequence, making them experience a proton ingestion event \citep{Arthur_PIE_2024A&A...691L...7C} or exotic binary interaction (e.g., \citealt{2013A&A...559A..66S}).

In this paper, we report the discovery of a LSP in the RV curves of two J-type carbon stars: VZ~Per and UV~Cam. From spectroscopic and photometric variability, we further constrain the asymmetric distribution of gas and dust around these systems, providing a unique testbed for investigating the origin of the LSP phenomenon.

The paper is structured as follows. In Sect.~\ref{sect:observation} we describe the spectroscopic data, in Sect.~\ref{sect:time-series} we assess the variability of the different time series datasets and derive orbital elements from the RV curve for the two stars. In Sect.~\ref{sect:gas}, we analyze their spectral line profile variation to evaluate the change in the gaseous component with the LSP phase. In Sect.~\ref{sect:dust}, we analyze the color variation of the VZ~Per light curve to constrain the dust mass and distribution within the system. In Sect.~\ref{sec:discussion}, we confront our results with the binary-induced model to explain the LSP and discuss the different interpretations. In Sect.~\ref{sect:conclu} we summarize our findings and provide a road for future investigation of both LSP variables and J-type star classes. 

\section{Observations} \label{sect:observation}

VZ Per and UV Cam were observed with the HERMES spectrograph mounted on the Mercator telescope \citep{HERMES} from 2009 to 2021. For VZ Per, a total of 97 spectra with a median exposure time of 1200s were collected. For UV Cam, 82 spectra were collected, with a median exposure time of 80s. The high-resolution mode was used, providing a resolution of $86\,000$ over the wavelength range $3800-9000$ Å. The dedicated Python-based pipeline was used for the data reduction (described in Sect. 3 of \citealt{HERMES} and Sect. 2.2 of \citealt{hermes_new}). The RV was obtained by cross-correlating the object spectrum with a carbon-star template and fitting a Gaussian function to the mean line profile, as represented by the resulting cross-correlation function (CCF). The mask used to compute the CCF is derived from a HERMES spectrum of the carbon star HD 7561 (Z Psc) observed on 4 September 2009. All lines with a depth below the relative level of 0.85 with respect to the local continuum were kept in the mask. A Doppler correction of $-$6.89 km/s was applied to set the mask at zero velocity.
The cross-correlation is discrete and determines the sum of all the fluxes of the wavelengths of all the lines in the mask over a range of velocities. The lines in the mask are shifted within a given range, and this results in a discrete cross-correlation profile when the sum of all the fluxes of the shifted wavelengths is plotted against the velocity shift. It is performed in two steps: the velocity is first estimated from a cross-correlation profile spanning a wide range of velocities. Then, the mask is shifted in discrete steps of 1 km/s over a range of $-50$ to $+50$ km/s around the estimated velocity. Finally, the resulting cross-correlation profile is fit with a Gaussian function to determine the radial velocity and its intrinsic uncertainty.
An uncertainty of 0.07 km/s expressing the long-term stability of the spectrograph is added quadratically to the formal uncertainty of the CCF fit to obtain the total uncertainty on the RV. This long-term accuracy is estimated from the average RV standard dispersion of a sample of RV standard stars monitored along with the science target (see Fig. 10 in \citealt{HERMES})\footnote{The RV precision was recently re-evaluated downward and now reaches 0.058 km/s \citep{hermes_new}.}. 
The log of RV observations is displayed in Tables~\ref{tab:RV_log} and~\ref{tab:RV_log_UVCam}. 

In addition to the spectroscopic observations, publicly available light curves are used in this work as complementary dataset for the analysis: the ASAS-SN photometry in the $g$ and $V$ bands\footnote{Available through the ASAS-SN sky Patrol query interface: \url{http://asas-sn.ifa.hawaii.edu/skypatrol/}} (see Sect. \ref{sect:time-series}) and the Gaia DR3 light curves in the $G_{BP}, G$, and $G_{RP}$ bands (see Sect.~\ref{sect:dust}). For both surveys, only the dataset for VZ Per was used, as UV Cam is too bright and exceeds the saturation limit in the ASAS-SN North detector ($g \leq 11.5$ mag, \citealt{ASAS_SN_2023arXiv230403791H}), and no photometric time series were available in the Gaia DR3 release.

\section{Characterization of the RV curve}\label{sect:time-series}

In this section, we analyze the variability of the RV curve for the two targets, determine the LSP, and model it as a possible signature of Keplerian motion induced by a low-mass companion.
\subsection{Period analysis of the RV and light curves}
\begin{figure*}[t]
    \centering
    \includegraphics[width=\linewidth]{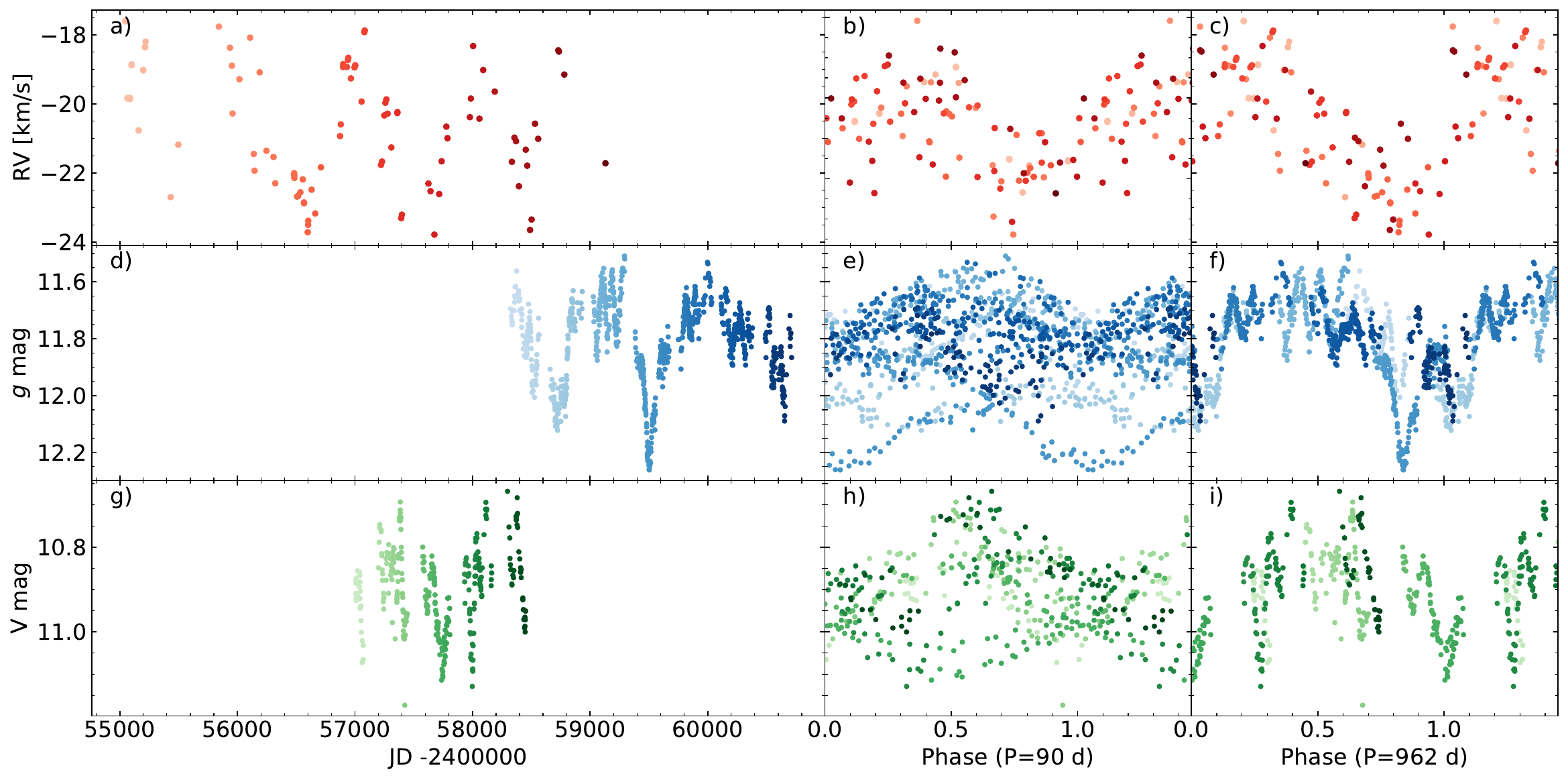}
    \caption{ RV and light curves of VZ Per. Top row: RV curve from HERMES. Middle row: $ g$-band photometry. Bottom row: $ V$-band photometry. Left column: time
series observations. Middle column: Data phase-folded on the 90-d period. Right column: Data phase-folded on the 962-d period. In panel b), the data plotted are the residuals after suppressing the LSP signal.}
    \label{fig:LC_VR_VZPer}
\end{figure*}

The RV curve of VZ Per is shown in Fig.~\ref{fig:LC_VR_VZPer}, together with the light curve in the $g$ and $V$ bands. Their respective periodograms are shown in Fig.~\ref{fig:Lombscargle}. Both RV and light curves reveal two variability periods: 1) At the longest periods, a peak at $\sim$962~d dominates the signal and can be seen as a sinusoidal variation in the RV curve with an amplitude of about 2~km/s, and regular dimming in the $g$ band filter. This will be referred to as the LSP. 2) At shorter periods, a $\sim 90$~d variation corresponding to the pulsation period. This is clearly seen in the light curve when phase-folded (see Fig.~\ref{fig:LC_VR_VZPer}, middle column) and detected in the RV curve periodogram when the LSP contribution is removed (see Figs.~\ref{fig:LC_VR_VZPer} and \ref{fig:Lombscargle}). 
Translating these variations in the PL diagram, the $\sim960$~d period lies on the D sequence, as expected for a LSP, while the $\sim 90$~d pulsation period lies on the C' sequence, implying a pulsation in the first overtone mode \citep{LSP_mass_loss_2019MNRAS.484.4678M}.

\begin{figure*}[t]
    \centering
    \includegraphics[width=\linewidth]{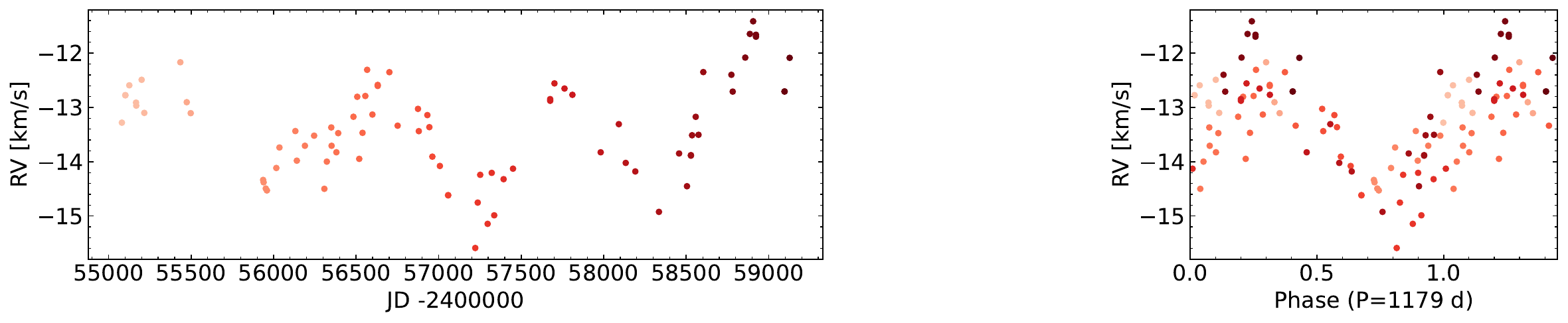}
    \caption{ RV curve of UV Cam. Left column: time
series observation. Right column:
Data phase-folded on the 1179-d period.}
    \label{fig:LC_VR_UVCam}
\end{figure*}

Similar conclusions can be reached for UV~Cam: the RV curve presents a clear modulation at $\sim1180$~d (see Figs.~\ref{fig:LC_VR_UVCam} and \ref{fig:Lombscargle}). In contrast to VZ~Per, no short-period signal, corresponding to intrinsic variability, is detectable in the RV curve, even after suppression of the long-period signal. The absence of photometric time series for this source prevents us from assessing whether the stellar variability is irregular or whether its brightness variability does not translate into an RV signal.

\subsection{Orbital parameters}

\begin{figure}[ht]
    \centering
    \includegraphics[width=\linewidth]{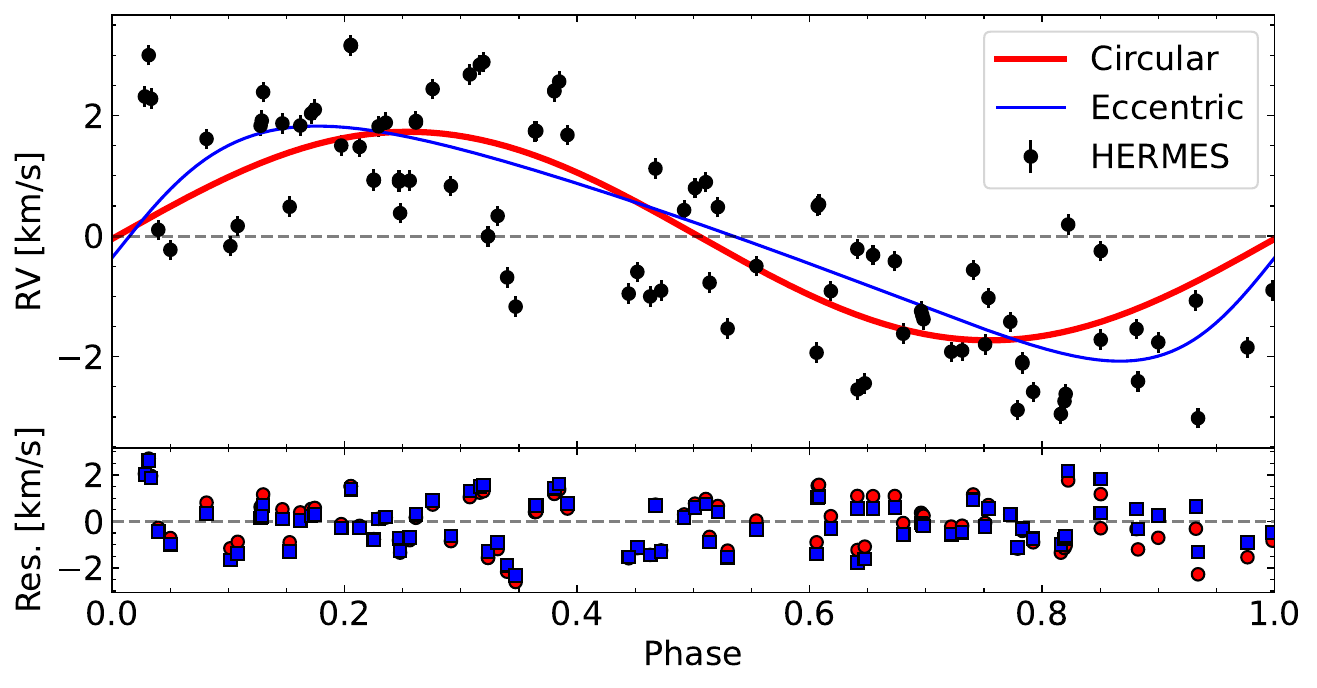}
    \includegraphics[width=\linewidth]{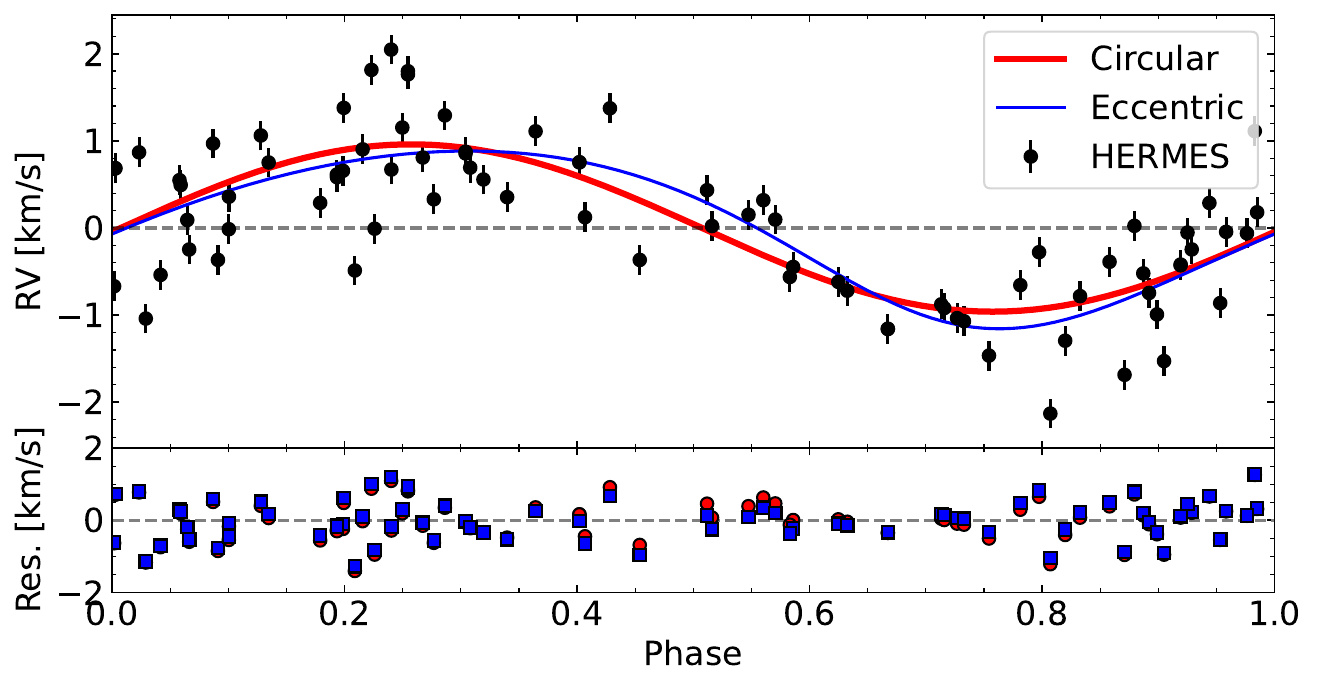}
    \caption{Phase-folded orbits for circular and eccentric solutions and respective residual values. Top: VZ~Per, bottom: UV~Cam. For the residuals, the red circles and the blue squares represent the residuals from the circular and eccentric solutions, respectively. Most of the symbols overlap.  }
    \label{fig:Orbit}
\end{figure}

The Keplerian orbit was retrieved from the RV curve using the \texttt{RadVel} software \citep{2018PASP..130d4504F}. 
For both stars, we fit a circular orbit as well as an eccentric one. The obtained parameters are listed in Table \ref{tab:orbital_parameters} and plotted in Fig.\,\ref{fig:Orbit}. Given the larger scatter arising from the intrinsic variability of the AGB star, it is difficult to place meaningful constraints on the eccentricity, even if the eccentric model has a lower Bayesian Information Criterion (BIC).

\begin{table}[t]
\caption{Orbital parameters derived from the RV curve}
    \centering
    \begin{tabular}{l|lrr}
    \hline
    \hline
        & & Circular & Eccentric \\
         \hline
       \parbox[t]{2mm}{\multirow{9}{*}{\rotatebox[origin=c]{90}{VZ Per}}} &$P$ [days] & 963.4$\pm$1.7& 964.6$\pm$1.4\\ 
        &$e$ &  0& 0.31$\pm$0.01\\
        &$\omega$ [$^\circ$]& $-$& 258$\pm$2\\
        &$T_C$ [MJD] & 54853& 54842\\
        &$K_1$ [km/s] & 1.73$\pm$0.02& 1.95$\pm$0.02\\
        &$V_\gamma$ [km/s] & $-$20.76$\pm$0.01&$-$20.78$\pm$0.01 \\
        &BIC & 8779& 8106\\
        
        &$a_1\sin(i)$ [au] & 0.153$\pm$0.002 & 0.164$\pm$0.03\\
        &$f$ [$10^{-4}\,\rm M_\odot$]&$5.1\pm0.2$ &   6.3$\pm$0.1\\
        \hline
         \parbox[t]{2mm}{\multirow{9}{*}{\rotatebox[origin=c]{90}{UV Cam}}}&$P$ [days]& 1177.9$\pm$3.4 & 1172.2$\pm$4.2\\
        &$e$ & 0& 0.15$\pm$0.02\\
        &$\omega$ [$^\circ$]& $-$& 153$\pm$8\\
        &$T_C$ [MJD] & 55107 & 55076\\
        &$K_1$ [km/s] &0.96$\pm$0.02 & 1.02$\pm$0.02\\
        &$V_\gamma$ [km/s] &$-$13.46$\pm$0.02 & $-$13.45$\pm$0.02\\
        &BIC & 2583& 2544\\
        &$a_1\sin(i)$ [au] & 0.104$\pm$0.002& 0.108$\pm$0.003\\
       &$f$ [$10^{-4}\,\rm M_\odot$]&$1.1\pm0.1$ & 1.2$\pm$0.1\\

\hline   
    \end{tabular}
    \tablefoot{$P$ is the orbital period, $e$ the eccentricity, $K_1$ the semi-amplitude of the giant's RV curve, $V_\gamma$ the center-of-mass velocity, $a_1$ the semi-major axis of the orbit of the giant around the center-of-mass of the system, $i$ the inclination, $f$ the mass function, $T_C$ is the epoch of inferior conjunction, $\omega$ is the argument of the periastron. 
      }

    \label{tab:orbital_parameters}
\end{table}

The absence of astrometric constraints prevents us from obtaining an estimate of the orbital inclinations. Because the parallaxes of both stars are close to or lower than 1.2 mas, the Hipparcos data are not precise enough to provide supplementary information (e.g., proper motion) that could deliver further constraints through their combination with more recent Gaia data. The absence or presence of astrometric signals in the closest LSP targets reported in Gaia DR3 is extensively discussed in 
 \cite{2026PASP..138e4202S} and \cite{2026arXiv260605292I}. 

Hence, without further constraints on the orbital motion in the plane of the sky, we cannot derive the orbital inclination and therefore can only derive a range of values for the companion mass and the orbital separation.
For $i= 30^{\rm o} \rightarrow 90 ^{\rm o}$, assuming a $1\rm \,M_\odot$ primary and the mass function derived from Table~\ref{tab:orbital_parameters}, the companion mass should range from 0.18 to 0.09  $\rm \,M_\odot$ for VZ Per and from 0.11 to 0.05  $\rm \,M_\odot$ for UV Cam.
Given the modulation present in the VZ Per light curve, the inclination can be further constrained to $i>60^{\rm o}$, leading to $M_2 < 0.1\, \rm M_\odot$.
For both systems, the companion nature is restricted to be at the brown dwarf limit, as commonly observed in the mass distribution of companions of LSP systems with available RV curves (\citealt{Nicholls_2009MNRAS.399.2063N}, \citealt{2026PASP..138e4202S}). 
Taking the projected semi-major axis and assuming $i\geq60^{\rm o}$, the Roche radius of the primary, setting an upper limit on the stellar radius to avoid Roche-lobe overflow, is $1.16\,$au for VZ Per and 1.4\,au for UV Cam, assuming  $M_1 =1\rm \,M_\odot$. 
Using the period-mass-radius relation for semi-regular variables pulsating on the first overtone mode from \cite{2019MNRAS.482..929T}, the derived radius of VZ~Per is 0.8\,au assuming a mass of 1$\,\rm M_\odot$ (or 0.9\,au assuming a mass of 1.5$\,\rm M_\odot$). Hence, no Roche lobe overflow is expected for the retrieved orbital configuration.

\section{Spectral line profile analysis} \label{sect:gas}
In this section, we analyze the variability of the spectral-line profiles of the two carbon stars. 
 
\begin{figure}[t]
    \centering
    \includegraphics[width=\linewidth]{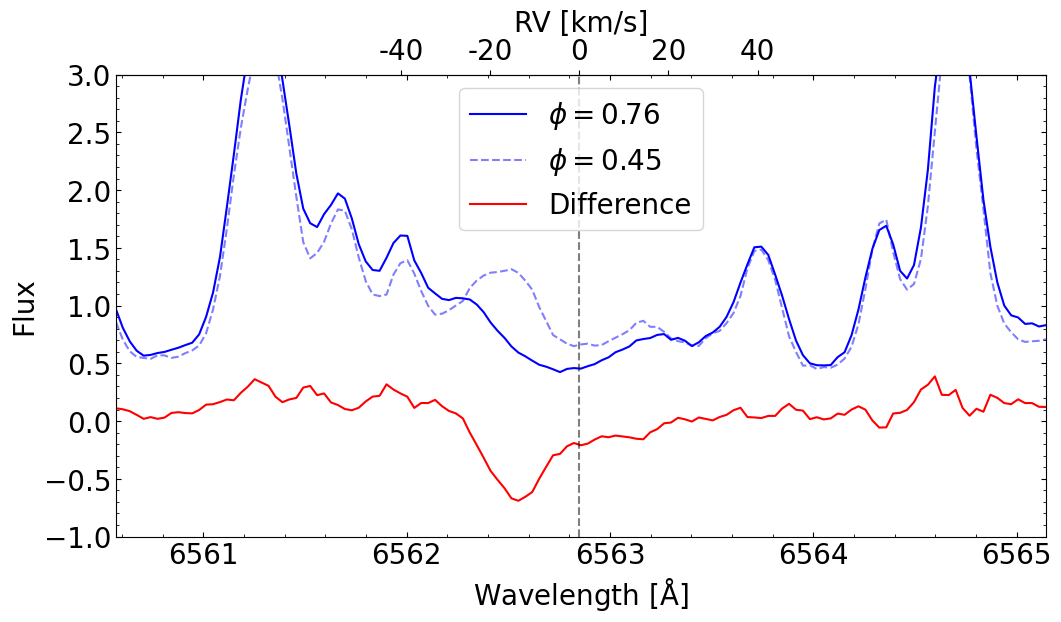}
    \caption{ Spectral profile around H$\alpha$ at two different phases for the star VZ~Per. The radial velocity zero-point (vertical dashed line) is set to the center-of-mass velocity listed in Table \ref{tab:orbital_parameters}.}
    \label{fig:difference_spectrum_Halpha}
\end{figure}

In the optical regime, bands from carbon-bearing molecules (such as CN and $\rm C_2$) dominate the spectrum of carbon-rich AGB stars. For small-amplitude pulsators (such as SRb or irregular variables), the relative intensity of absorption and pseudo-emission is not expected to vary significantly or periodically with time. 
However, for the two stars of interest, some atomic transitions exhibit amplitudes that vary in phase with their LSP derived from their RV curve. 
More specifically, both Balmer transitions, H$\alpha$ and H$\beta$, exhibit variable absorption profiles\footnote{Bluer Balmer transitions could not be detected due to the low S/N ratio in the blue part of the optical spectrum.}. 
Figure~\ref{fig:difference_spectrum_Halpha} illustrates the line-profile change between two spectra taken at different phases for the star VZ~Per. As the optical spectrum is crowded with molecular absorption, the continuum cannot be defined. Hence, in the figures presented here we set the pseudo-continuum-
normalized fluxes taken as the median value of the displayed spectral
window ( |RV| $\leq 60$ km/s). A selection of spectral profiles taken at different orbital phases is displayed in Appendix \ref{sect:spectral_line_overimposed}.
While pulsating carbon stars (Mira or SRa) can exhibit an emission component in the Balmer lines (strongest in H$\delta$) resulting from shock propagation (e.g. \citealt{2000A&A...362..655A}), strong  Balmer lines in absorption are not expected in cool carbon stars. 
H$\alpha$ absorption in the star UV Cam was previously reported by \cite{Barnbaum_C_stars_1994ApJS...90..317B}, together with the other J-type star TO Ser. But the variability was not assessed. From a comparison with other J-type stars, it seems that Balmer absorption is indeed specific to this spectral type and that the absorption depth can vary over time (see Appendix~\ref{sect:others_J-type_stars}).

\begin{figure}[t]
    \centering
    \includegraphics[width=\linewidth]{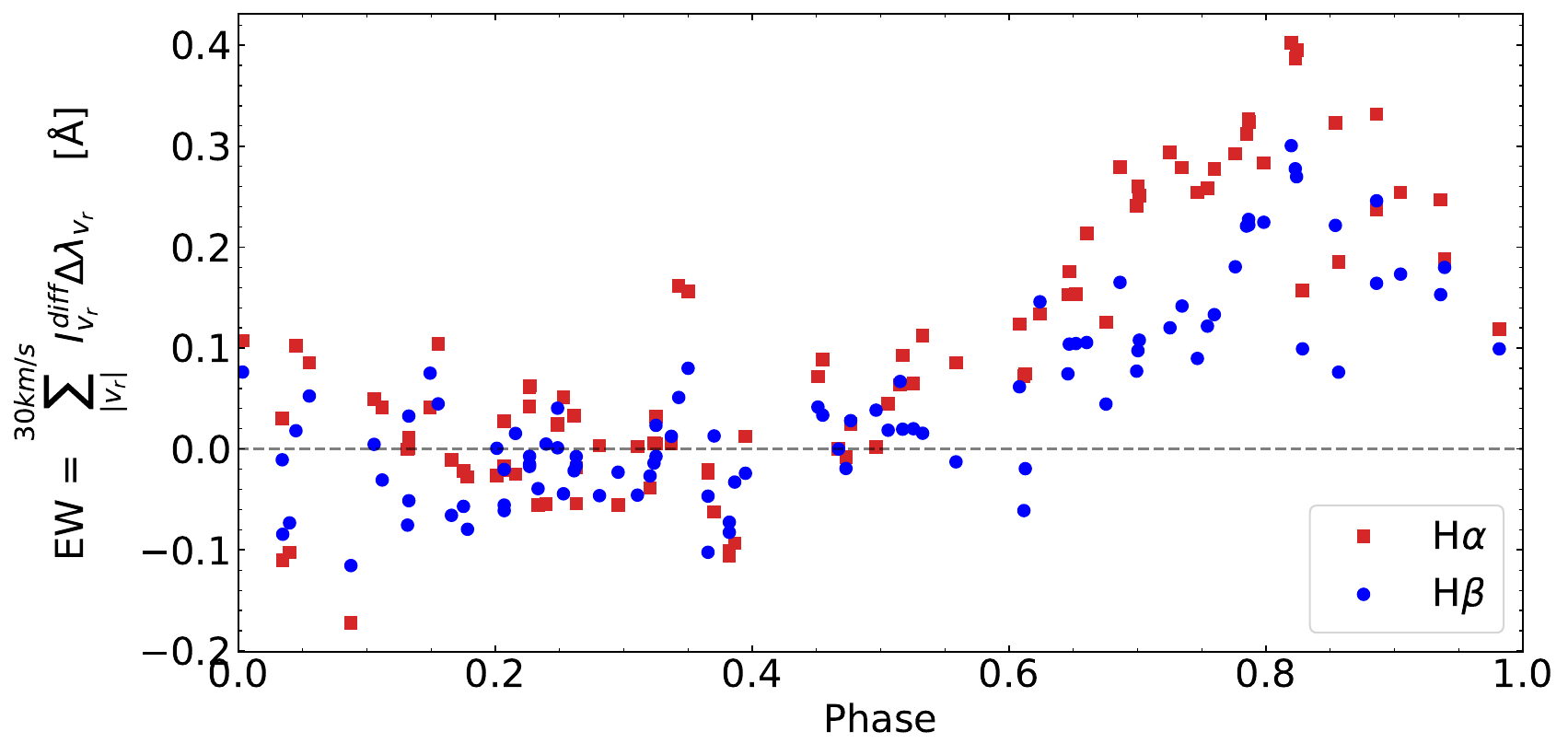}
    \includegraphics[width=\linewidth]{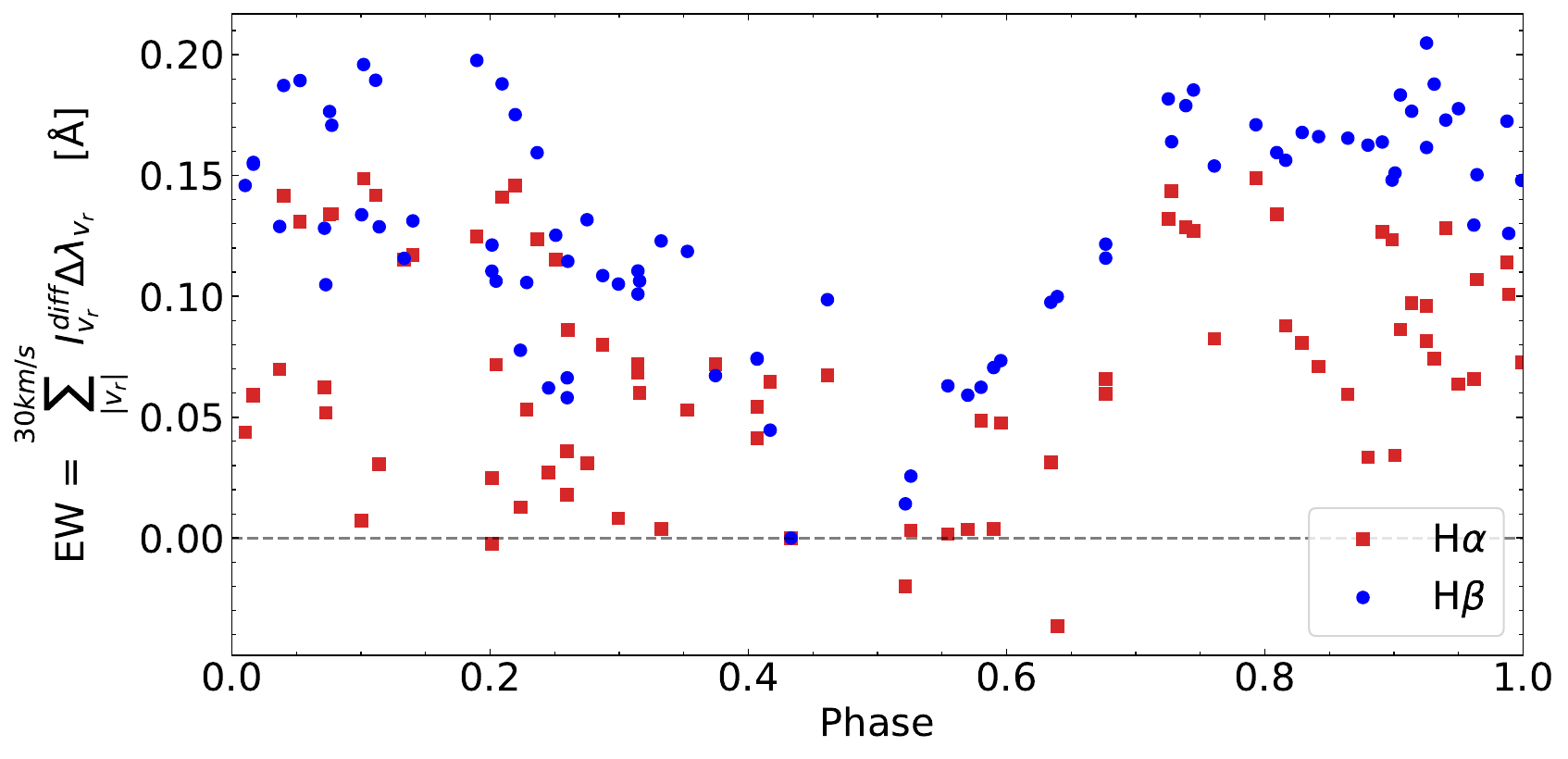}
    \caption{Equivalent width of the Balmer H$\alpha$ and H$\beta$ transitions as a function of orbital phase. Top panel: VZ Per. Bottom panel: UV Cam.}
    \label{fig:EW}
\end{figure}

To assess the change in profile only, and not the displacement in velocity of the photospheric contribution, we also analyzed the difference spectra. They are defined as the difference between the spectrum at a specific phase and a reference spectrum taken at the phase where the absorption is shallowest (at $\phi \sim 0.4$ for both stars; Fig.~\ref{fig:difference_spectrum_Halpha}). 
The orbital-phase dependence of the H$\alpha$ and H$\beta$ absorption features in the two stars is illustrated in Fig.~\ref{fig:EW}, where the equivalent width (EW) of the difference spectra is plotted against the phase for VZ~Per and UV~Cam. 
For VZ~Per, the absorption increases in the phase interval 0.5 to 1.0, reaching its maximum value around 0.8. For  UV~Cam, the variability is less pronounced but exhibits larger values from phases 0.6 to 1.3 (or, conversely, a shallow absorption is present between phases 0.3 and 0.6). 
A correlation between the H$\alpha$ equivalent width and the long period of LSP O-rich variables was previously mentioned by \cite{Wood_LSP_spectro_2004ApJ...604..800W}, who interpret this as a potential signature of chromospheric activity. But chromospheric evidence is not expected to be detectable in the optical range for C-rich (hence cooler) stars due to molecular opacity and should only manifest as UV emission lines \citep{1989ApJ...345..543L}. Hence, the detection of H$\alpha$ variability in the spectrum of C-rich stars with a LSP challenges the chromospheric origin to explain the H$\alpha$ variability.

It can be seen that the phase behavior slightly differs between the two stars. Given that the EW variation is more pronounced in VZ Per, where the RV variations are about twice larger than in UV\,Cam, one could speculate that there is a correlation between the amplitude of the RV curve and the strength of the absorption lines. But without any light curve for UV\,Cam, we cannot conclude whether this variation also translates into the shape of the LSP photometric modulation.

\begin{figure}[th]
    \centering
    \includegraphics[width=0.495\linewidth]{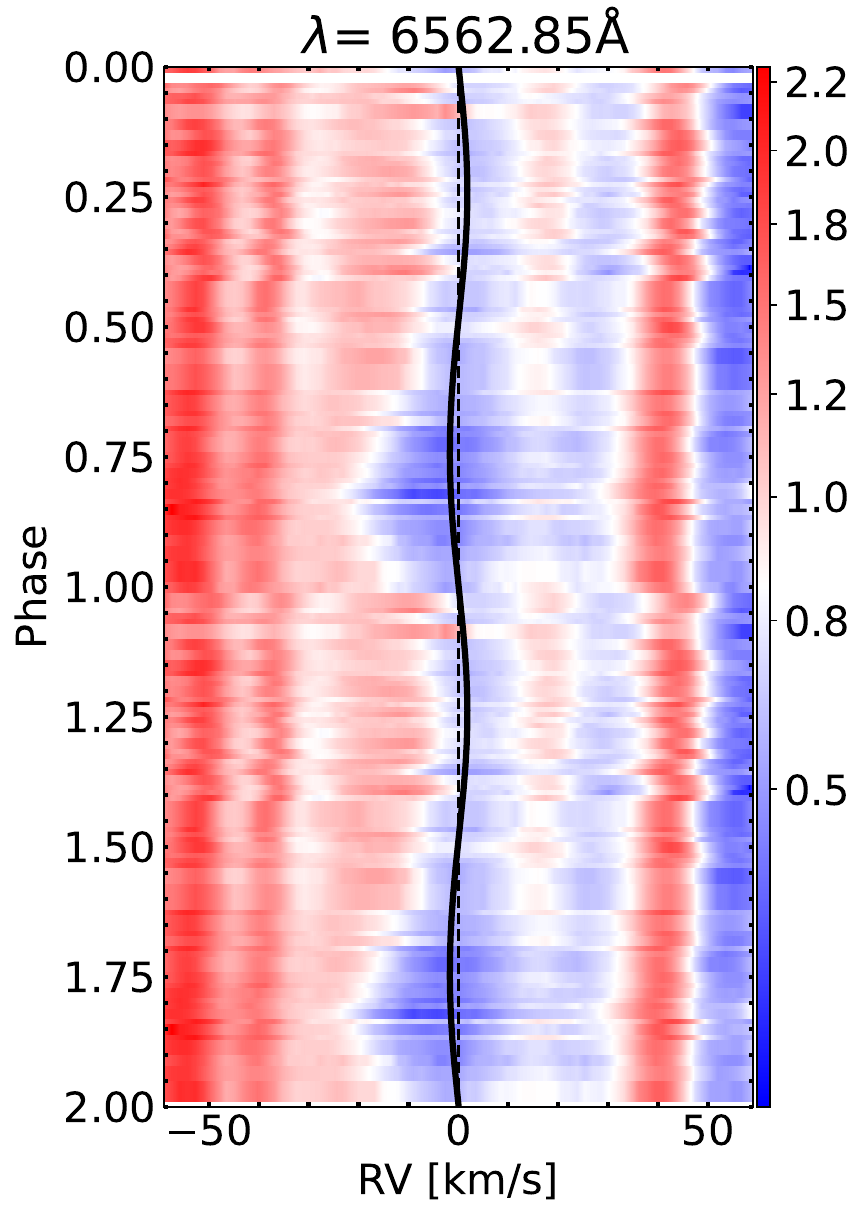}
    \includegraphics[width=0.495\linewidth]{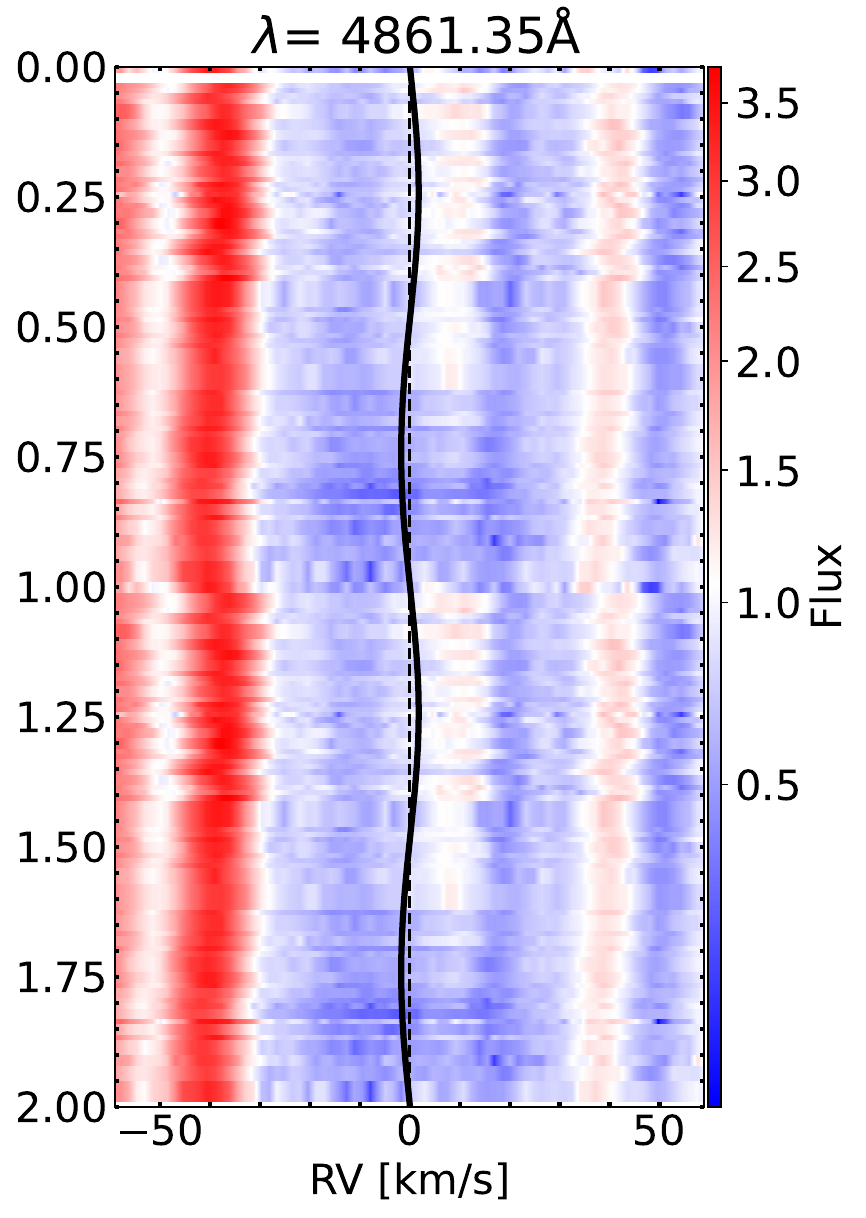}
        \includegraphics[width=0.495\linewidth]{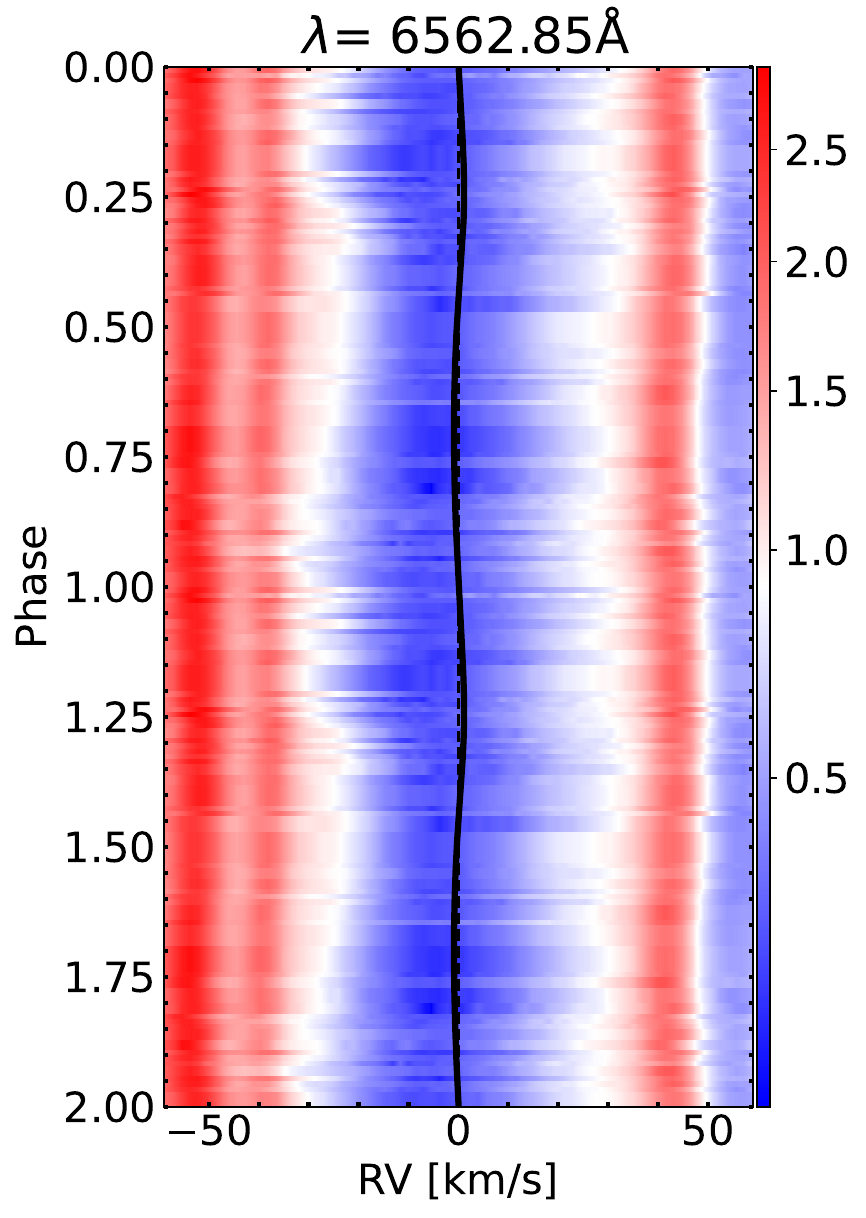}
    \includegraphics[width=0.495\linewidth]{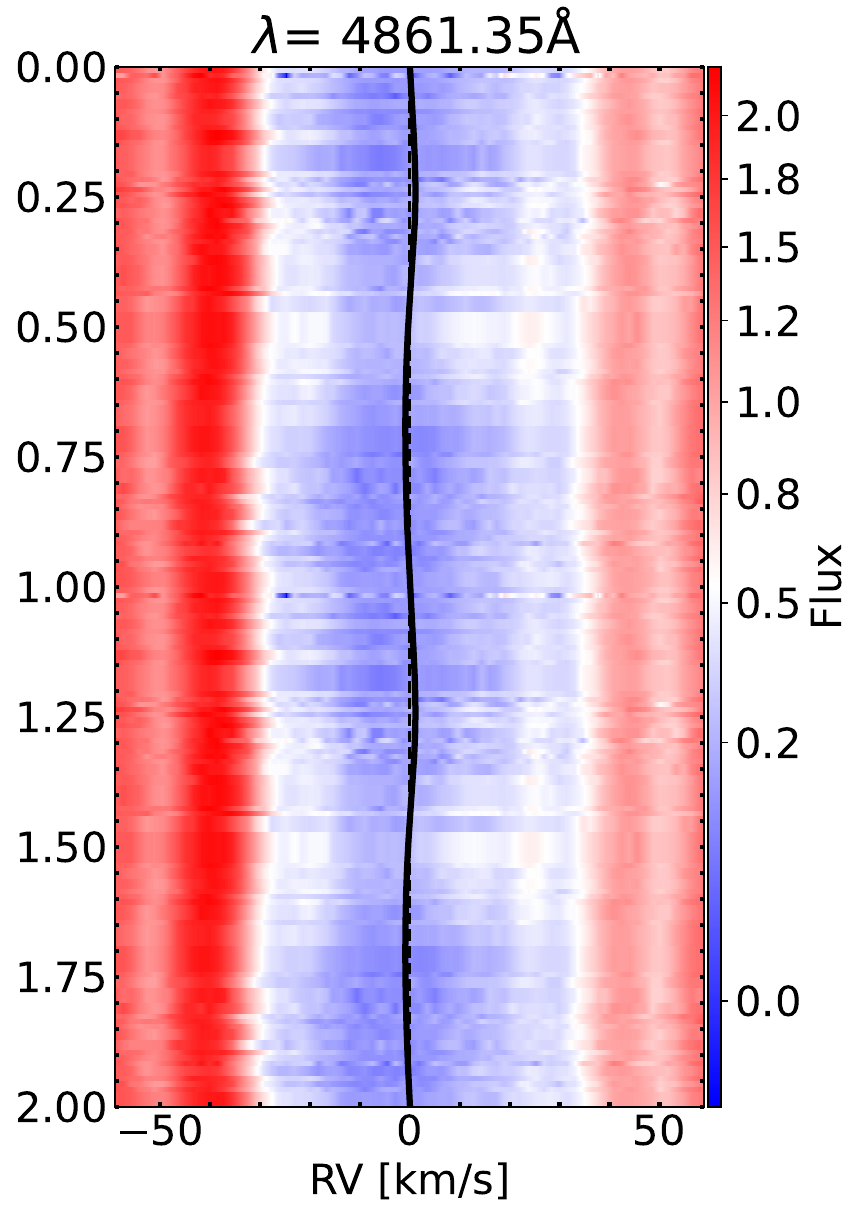}
    \caption{Dynamic spectra of the H$\alpha$ (left panels) and H$\beta$ (right panels) lines as a function of the orbital phase. Top panels: VZ Per. Bottom panels: UV Cam. In each panel, the observed phases are shown twice to guide the eyes; time runs down. The color scale (see the color map displayed along the right-hand side axis of each panel) represents the pseudo-continuum-normalized fluxes, taken as the median value over the spectral window. The solid line corresponds to the RV variation associated with the LSP, while the vertical line marks the systemic velocity, and the horizontal lines on the left indicate the observed phases. }
    \label{fig:DynamicSpectra_VZPer}
\end{figure}

\begin{figure}[t]
    \includegraphics[width=0.49\linewidth]{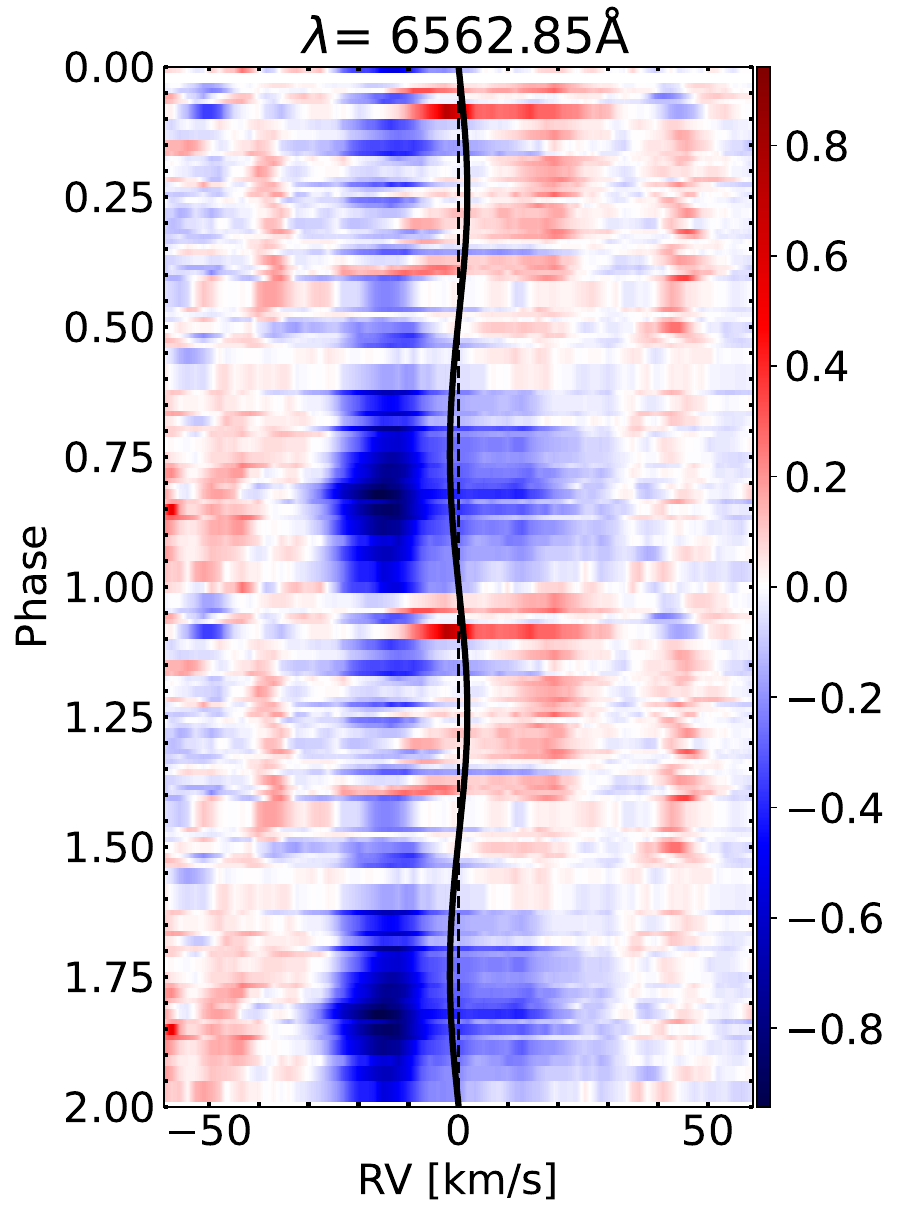}
    \includegraphics[width=0.49\linewidth]{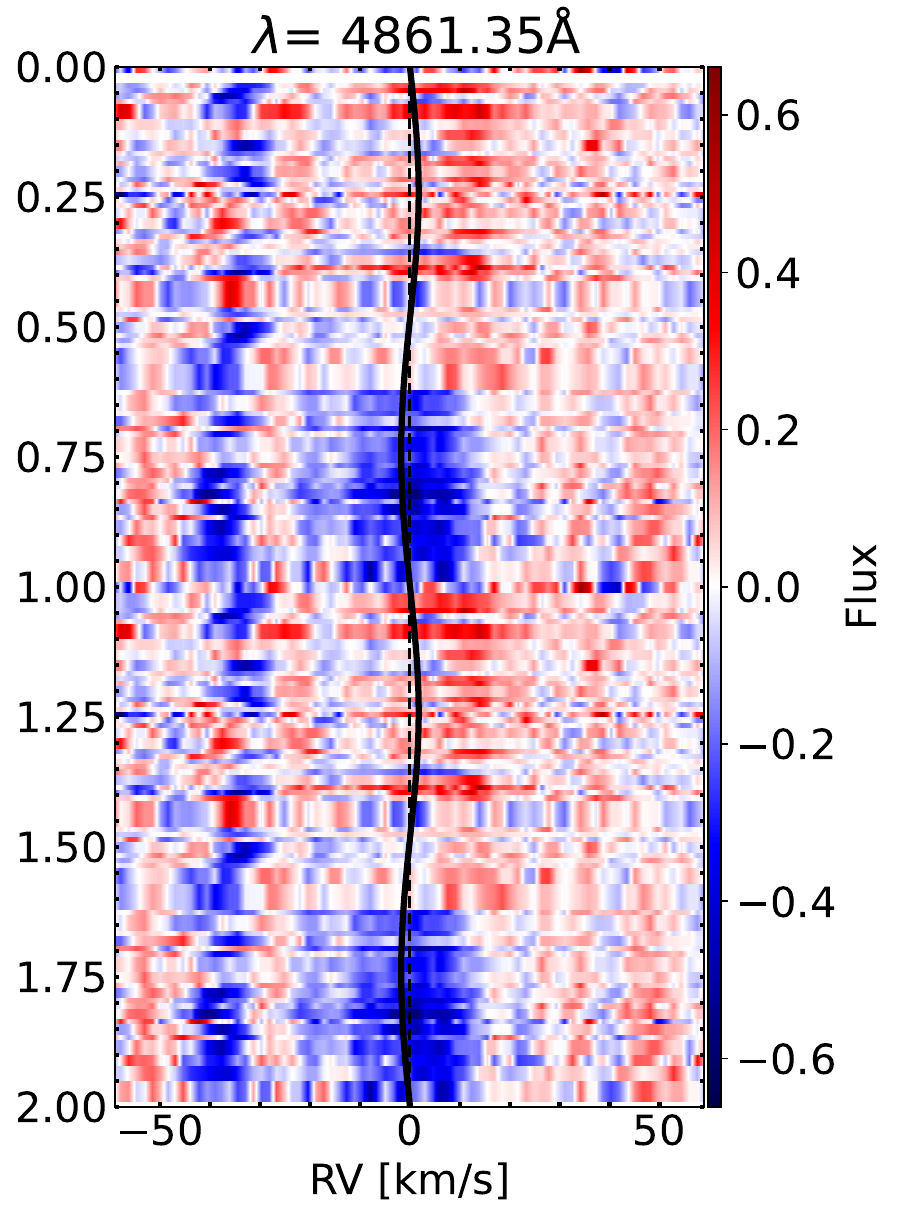}

    \includegraphics[width=0.49\linewidth]{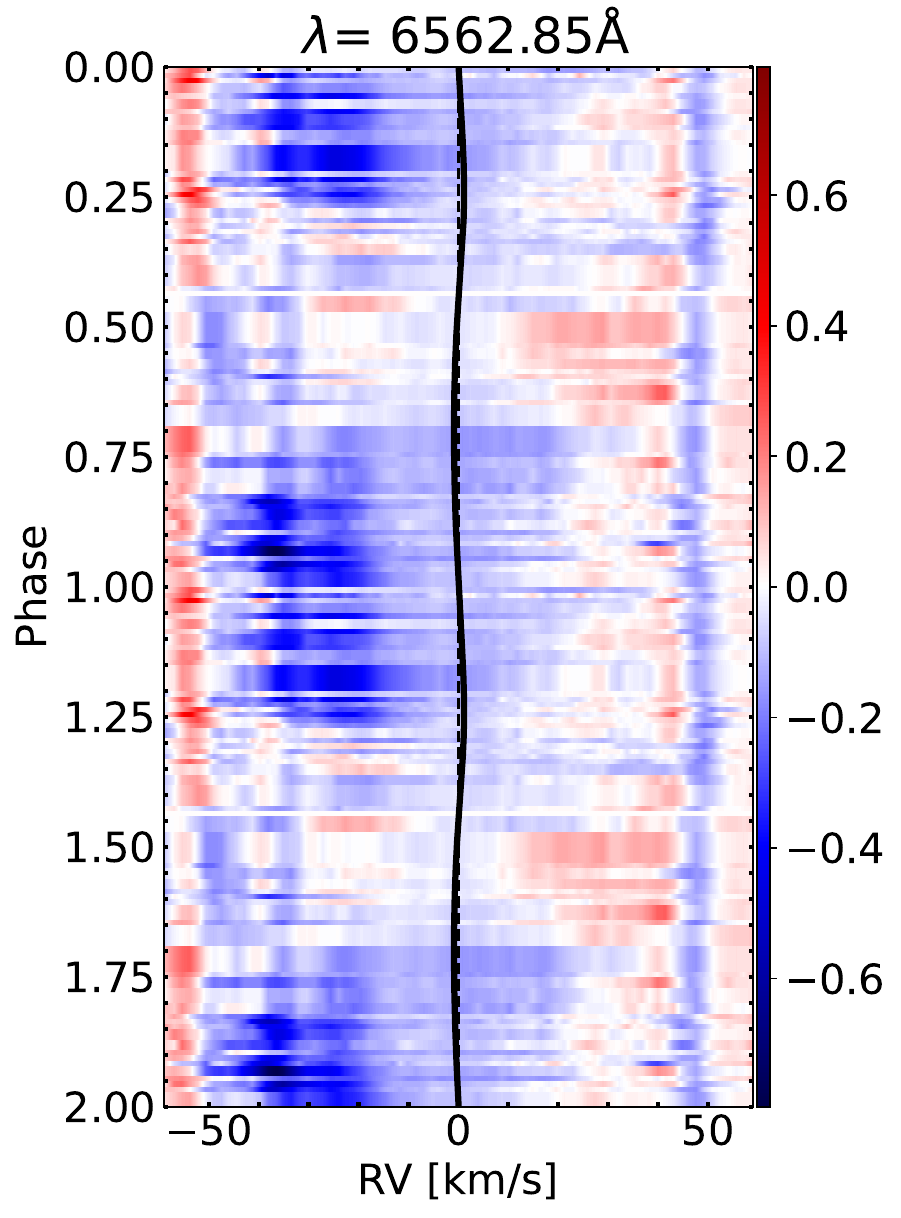}
    \includegraphics[width=0.49\linewidth]{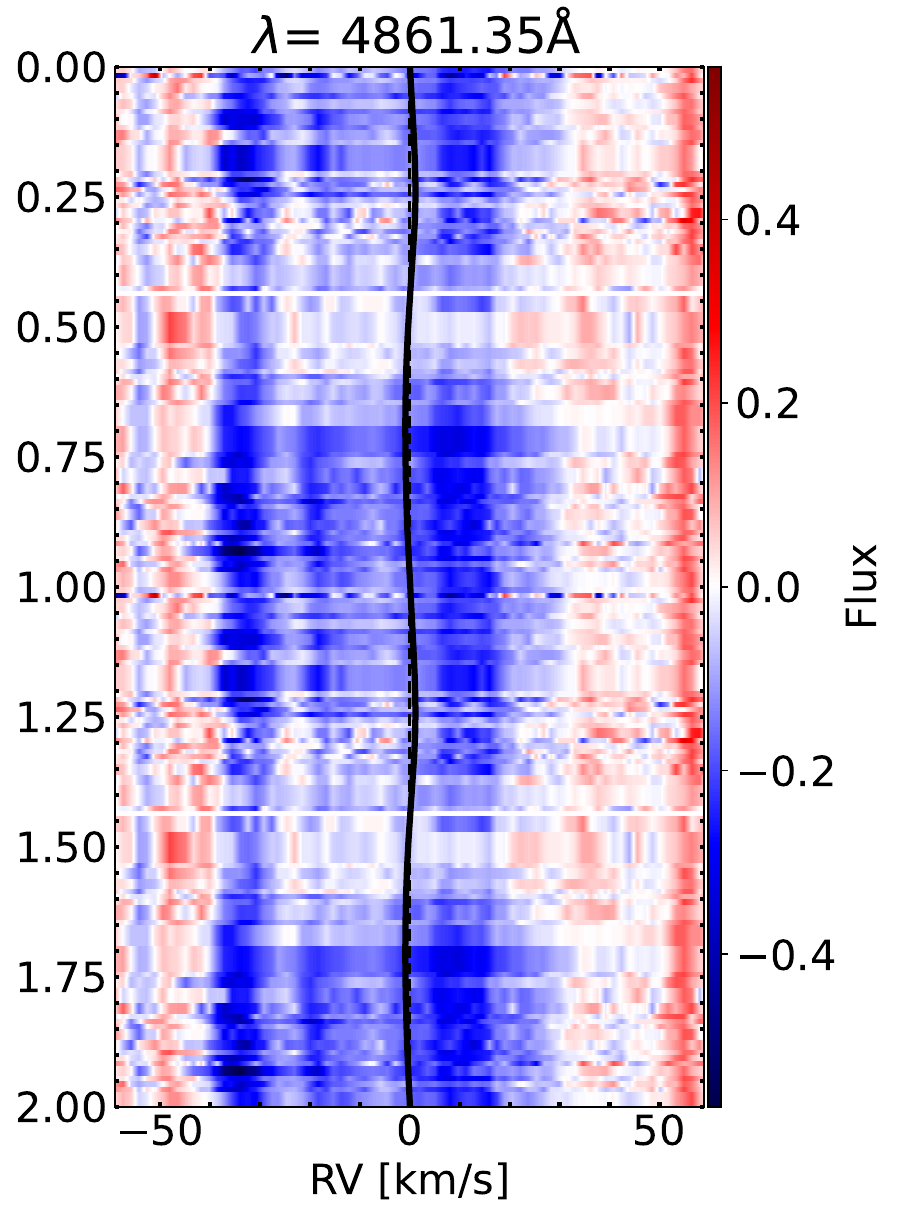}
    \caption{Dynamic difference spectra of the H$\alpha$ (left panels) and H$\beta$ (right panels) lines as a function of the orbital phase. Bottom row: UV Cam, Top row: VZ Per. The color represents the residual fluxes after subtraction of a model spectrum.  } 
    \label{fig:dynamic_spectrun_diff}
\end{figure}

To highlight the shape change of the absorption feature with phase, we present the dynamic spectrum for both stars and both Balmer transitions in Fig.~\ref{fig:DynamicSpectra_VZPer} and for their corresponding difference spectrum in Fig~\ref{fig:dynamic_spectrun_diff}. In a dynamic spectrum, the change of the radial velocity ($x$-axis) and intensity (color-coded) of a specific line is plotted as a function of the LSP phase ($y$-axis, top to bottom). The LSP cycle is displayed twice to guide the eye. 
This visualization emphasizes smooth but faint changes in the spectral profile, which are not immediately obvious when only looking at superimposed spectral line profiles, as displayed in Figs.~\ref{fig:spectra_overimposedVZPer} and \ref{fig:spectra_overimposedUVCam}. To enhance the low-flux details, the colorbar in Fig.\,\ref{fig:DynamicSpectra_VZPer} follows the square root of the flux. When removing the photospheric contribution through the computation of the difference spectrum, as shown in Fig.\,\ref{fig:dynamic_spectrun_diff}, the flux variations appear more clearly.
The strengthening in absorption is seen in these images as the blue band, whose vertical extent defines the phase coverage, and whose horizontal extent corresponds to the radial-velocity width of the absorption feature, here ranging up to several tens of km/s, about an order of magnitude larger than the RV-curve amplitude.

\section{Multi-band photometric analysis} \label{sect:dust}

In this section, we assess the color variation associated with the LSP in VZ Per and infer the corresponding dust properties required to explain the dimming event as due to obscuration by a moving dusty structure.

\subsection{Gaia light curves}
\begin{figure}[t]
    \centering
    \includegraphics[width=\linewidth]{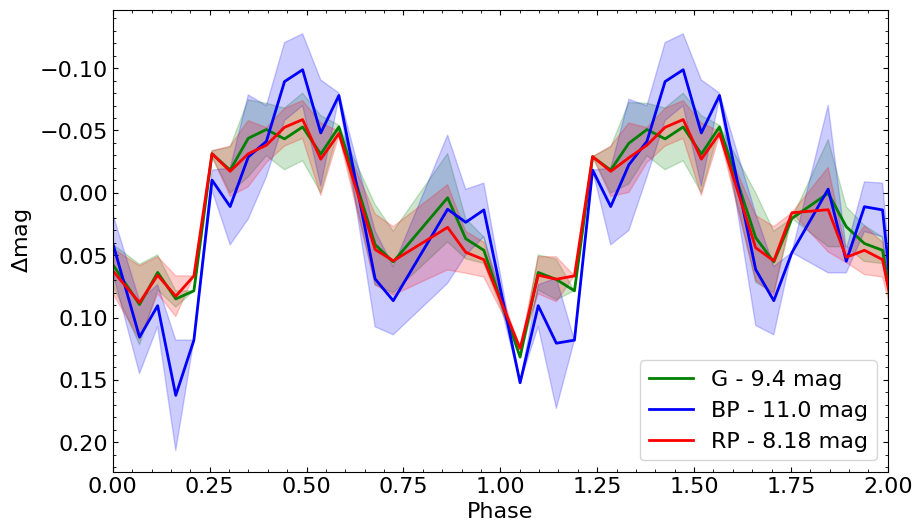}
    \caption{Smoothed light curves from the three Gaia passbands phase-folded on the LSP period. Each curve was obtained after removing the median and performing a rolling mean on a 90-day period.}
    \label{fig:gaia_rolling_mean}
\end{figure}

For VZ Per, light curves in different photometric bands are available. In particular, the Gaia DR3 light curve in the $G_{BP}$, $G$, and $G_{RP}$ filters spans about 34 months, covering the 962-day period of the LSP and allowing us to assess its color dependence. 
To deduce the magnitude variation over the LSP only and remove the short-term variability due to the pulsation, we compute a rolling mean over 90 days as shown in Fig.~\ref{fig:gaia_rolling_mean}, where the data, phase-folded, are displayed twice to avoid edge effects when computing the mean. 

The magnitude drop (difference between the maximum and minimum values) in each photometric band defines the attenuation at the corresponding wavelength. From Fig.~\ref{fig:gaia_rolling_mean}, we get $A_\lambda=$ (0.26, 0.19, 0.18)~mag for $\lambda_{\mathrm {ref}}$ = (532, 673, 797)~nm respectively. The wavelength dependence of the magnitude drop can be assessed by fitting a linear function to the wavelength variation of the attenuation on a logarithmic scale ($A_\lambda \propto \lambda^p$). We obtain $p = -0.9\pm0.2$, consistent with extinction by a dust particle in the Mie-diffusion regime ($p=-1$). 

\subsection{Dust obscuration}

Assuming the dimming event is due to obscuration by dust grains in the line of sight, we can compute the corresponding amount of dust required.

In carbon-rich AGB stars, the dust mineralogy is mainly composed of amorphous carbon grains. Assuming the usual grain size distribution ($n(a) \propto a^{-3.5}$, \citealt{MNR_1977ApJ...217..425M}) with grain radii $a$, ranging from $a_m= 0.005$ to $a_M = 0.1 ~\mu$m, the attenuation at wavelength $\lambda$ is given by:
\begin{equation}
\label{eq:dust}
 A_\lambda\; [\mathrm{mag}] = 1.086\;\pi\; N \;\frac{\int^{a_M}_{a_m} a^2 Q_{\lambda}(a)\;n(a)\;\mathrm{d}a}{\int^{a_M}_{a_m}\; n(a)\; \mathrm{d}a}
\end{equation}
with $N$ the column density of dust grains and $Q_\lambda$ the extinction efficiency of the dust particle. To evaluate $Q_\lambda$ we use the refractive index of amorphous carbon from \cite{1998A&A...332..291J}\footnote{It has to be noted that variations exist in the reported dust optical data and it can influence the derived observables; see, for example, \cite{carbon_dust_grains_2025A&A...697A..52S}.} under the Mie theory for spherical particles through the \textsc{miepython} package \citep{prahl_miepython_2026}. With the adopted dust distribution and Eq.~\ref{eq:dust}, the magnitude drop of 0.19 in the $G$ band requires a dust column density of $(2.0\pm 0.6)\times 10^{11}\,\rm cm^{-2}$. Taking a density of $2\, \rm g \,cm^{-3}$ for amorphous carbon grains \citep{AmC_1991ApJ...377..526R}, the corresponding mass column density is $(4.5\pm 1.2)\times10^{-6} \, \rm g \,cm^{-2}$. 

\begin{figure}[ht]
    \centering
    \includegraphics[width=\linewidth]{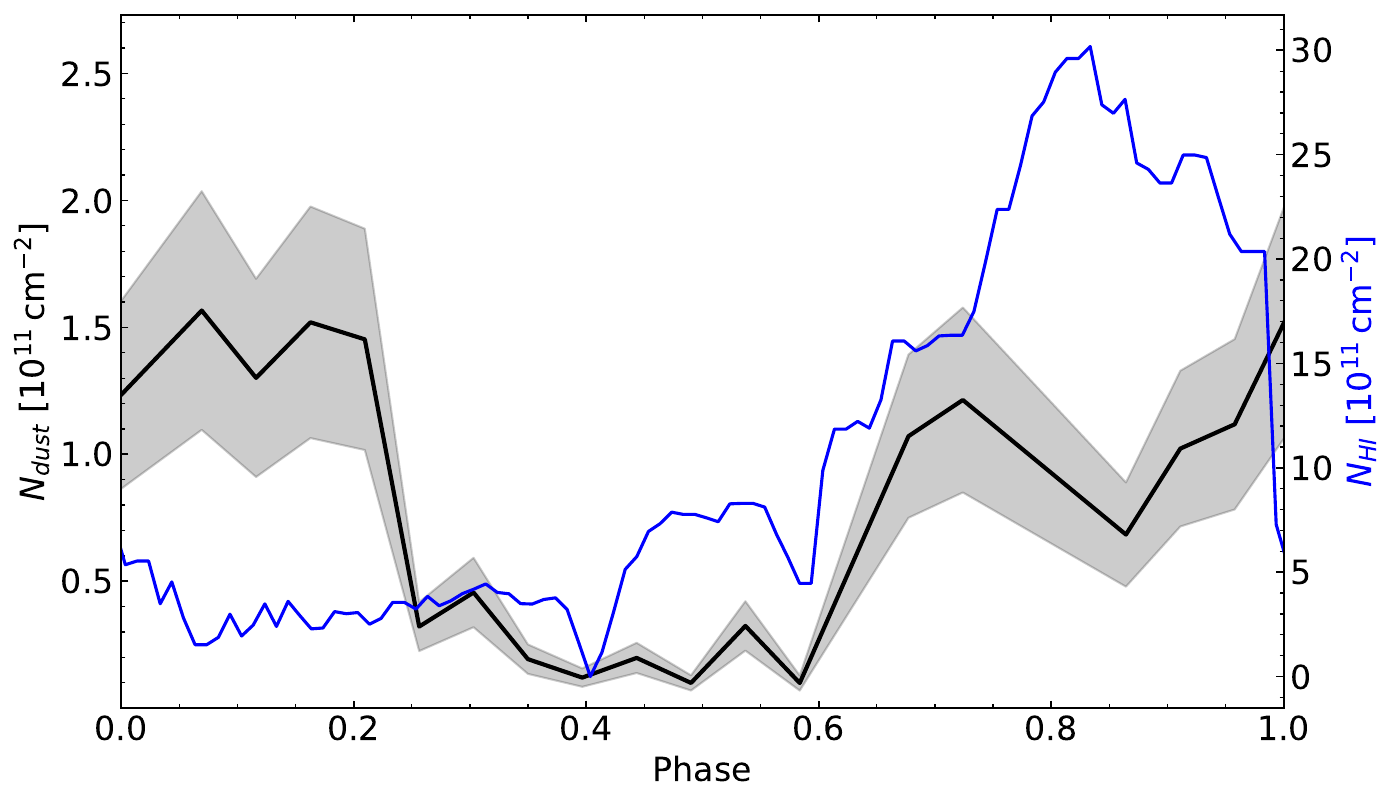}
    \caption{Dust column density (black curve) and lower limit on neutral hydrogen column density (blue curve) as a function of the LSP phase for VZ Per. }
    \label{fig:column_density}
\end{figure}

Estimating the surface of the obscuring body, we can obtain a derivation of the total dust mass in front of the line of sight. 
Without making any assumption on the geometry of the obscuring body, its surface can be bounded between the minimal surface assuming completely opaque grains and the stellar surface itself. A dimming of about 0.25 mag implies a drop of about 20\% of the stellar flux. The minimal occultation surface is therefore $0.2\pi\; R_*^2$, corresponding to a situation where 20\% of the stellar disc does not emit light (completely absorbed by dust). 
Taking $\rm R_*\sim200\, R_\odot$, the total dust mass of the minimal occultation surface would then be $\approx 3.5\times10^{-13} \rm \, M_\odot$, and the maximal surface, five times higher. 
As $N$ scales linearly with $A_\lambda$, the phase behavior displayed in Fig.\,\ref{fig:gaia_rolling_mean} can be translated into a phase-dependent variation of $N$. This is done in Fig.\,\ref{fig:column_density}, where the phase-dependent behavior of the absorption lines derived from Sect.\, \ref{sect:gas} is also plotted to compare the gas and dust variations. The column density of neutral hydrogen on the first excited level $N_{HI}$ is derived from the H$\beta$ equivalent width (EW) assuming an optically thin medium. As this assumption cannot be tested, the retrieved value should be treated only as a strict lower limit.

\section{Discussion}\label{sec:discussion}
In this section, we first interpret the results within the framework of a binary system in which a trailing spiral from a brown dwarf companion shapes the gas and dust. We then discuss constraints on the intrinsic variability model and, finally, the open questions regarding J-type carbon stars in light of the discovery of a LSP.
\subsection{Confrontation with the binary scenario}
In the previous section, we derived the orbital configuration and determined the column densities of gas and dust as a function of the LSP phase. These impose constraints on the underlying physical parameters (density, geometry, temperature, and composition) of the binary-induced wake. 

\begin{figure}[ht]
    \centering
    \includegraphics[width=\linewidth]{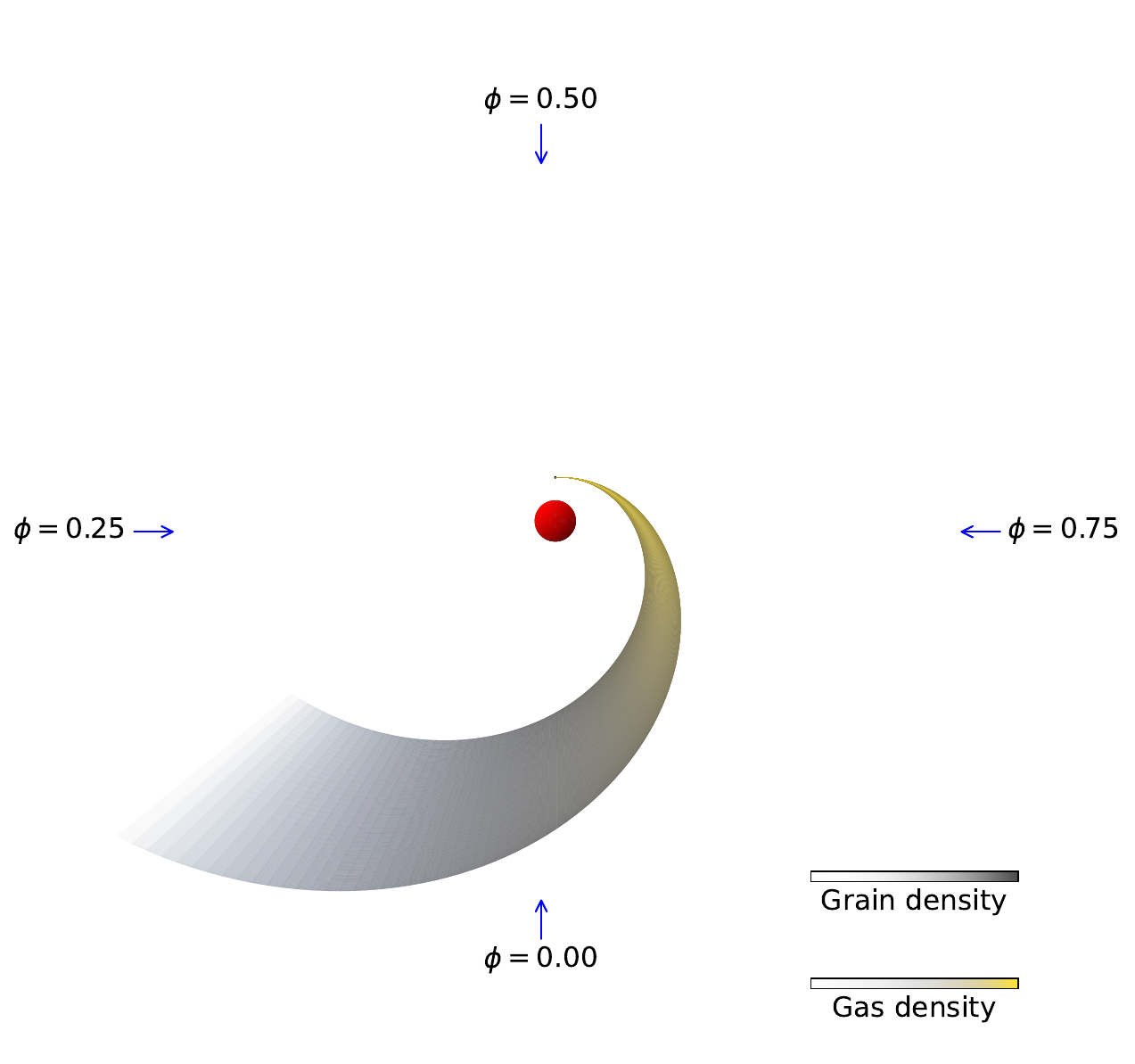}
    \caption{Sketch highlighting the link between the orbital phase and the viewing angle of the wake trailing a companion according to the binary scenario. }
    \label{fig:sketch_binary}
\end{figure}

For example, the observed Balmer absorption is naturally explained as the scattering of continuum photons by H atoms in the trailing, expanding wake of the companion. Due to the binary motion, the system is observed from different viewing angles, providing a tomographic view of the wake.
In this picture, Fig.~\ref{fig:column_density} can be interpreted as the gas -- probed by the Balmer line variations -- and dust -- probed by the multi-color light curve -- at different vantage points within the spiral wake trailing the sub-stellar companion. This is illustrated in Fig.~\ref{fig:sketch_binary}.
Assuming that the trailing wake created behind the companion follows an Archimedean spiral, its shape can be parameterized from the derived orbital configuration. The opening angle of the cone is set by $\alpha \approx \mathcal{M}^{-1} = c_s/v_{\mathrm{orb}} \approx $ 8 degrees, taking a sound velocity $c_s \sim 3$~km/s\footnote{The sound speed can be evaluated using $c_s = \sqrt{\frac{kT}{\mu_m m_H }} = 0.0912\sqrt{T} \rm ~km/s$ for H atoms. For circumstellar temperatures ranging from 1000~K to 1500~K, $c_s\in [3.0, 3.5]$ km/s. } and an orbital velocity of $\sim 20$ km/s (for $q = M_1/M_2 \approx 11$, in the edge-on configuration). In such a toy-model view of the wake, the portion of the stellar surface hidden behind the spiral is nearly 50\% at phase 0.8. 

EW variations of atomic transitions -- like Mn I -- with the phase of the LSP were recently detected for the LSP of Betelgeuse and interpreted as the imprint of a trailing wake caused by a companion embedded in the stellar wind of the supergiant \citep{2026ApJ...998...50D}.
On the other side of the companion-mass spectrum, dusty comet-like tails have also been reported for disintegrating exoplanets orbiting dwarf stars via asymmetric transit detections \citep{2025ApJ...984L...3H}.

\subsection{Confrontation with the pulsation scenario}
While the results reported here are consistent with the presence of a substellar companion to account for the LSP, it should be noted that other hypotheses exist to explain the origin of LSP in general (see Sect.\,\ref{sect:intro}). Here, we will briefly translate our observations into constraints on a potential intrinsic variability. 

The RV amplitude of the LSP in VZ~Per is about ~1.7 km/s (see Table \ref{tab:orbital_parameters}) and follows a sinusoidal curve. If linked to radius change, the semi-amplitude should be $\Delta R = \Delta RV \times P /(2\pi) \approx 0.16 \rm ~au = 34~R_\odot$.
Using UV Cam parameters, it corresponds to $0.12 \rm ~au$. It translates to a variation of about $\pm15\%$ of the stellar radius, where the largest expansion would occur at $\phi=0.75-0.80$, associated with the light curve pre-minimum. 
However, LSP variables cannot be associated with a radial pulsation as their period lies above the fundamental mode of radial pulsation (associated with Mira stars, lying on the sequence C, see e.g. \citealt{2017ApJ...847..139T}).

These considerations had led to interpret the radial velocity change in the framework of non-radial pulsation instead, such as a dipole oscillatory convective mode \citep{LSP_dipole_mode_2020MNRAS.492.1348T}, where the decrease of bolometric luminosity is mainly associated with a temperature change ($\Delta T \leq150$~K for carbon stars) leads to radius amplitude variation of $\Delta R \approx 12 \rm \, R_\odot$ (see also Fig.~5 in \citealt{LSP_dipole_oscillation_2015MNRAS.452.3863S}). If we consider the LSP as a luminosity variation phenomenon that correlates with an effective temperature change, then the observed increase of Balmer EW around the light minimum ($\phi \sim 0.8 $) should be interpreted as the spectral footprint of the corresponding temperature decrease. Such a temperature decrease should then also be responsible for color index variation and associated increase in dust density. 
 Since these models reproduce different phase shifts between RV and light curves as well as their respective amplitudes (see Fig. 8 in \citealt{LSP_dipole_mode_2020MNRAS.492.1348T}), they are compatible to some extent with the observation we report, even though they do not provide sufficient constraints to further test the viability of the scenario. Models tailored to specific stellar parameters, together with synthetic spectra (especially around H$\alpha$), would be required to refute or support this alternative hypothesis. In particular, given that these non-radial pulsations give rise to different observables depending on the star chemistry, such as phase shift in near-IR light curve due to different molecular opacity \citep{LSP_dipole_mode_2020MNRAS.492.1348T}, a search for spectroscopic differences between C-rich and O-rich LSP should be performed.
\subsection{J-type carbon stars connection}

The discovery of (substellar) companions around J-type stars is to be put in perspective with the question of an eventual link between binarity and the J-type peculiar spectral type.  
Binarity has been previously detected and characterized in the J-type star BM~Gem (\citealt{1991A&A...251...79B}, \citealt{Izumiura_2008}). But BM~Gem is part of the subclass of silicate-rich J-type stars (representing about $\sim 20$\% of J-type stars \citealt{Jstqrs_2007AJ....134..214C}) where their silicate-rich mineralogy around a carbon-rich AGB star is explained by the presence of a circumbinary or circumsecondary disc created through interaction with a stellar companion (\citealt{1990AJ.....99.1612L}, \citealt{Jstar_silicate_MIDI_2007A&A...467.1093D}). However, neither VZ~Per nor UV~Cam has been reported as exhibiting a silicate emission feature.

The J-type class of carbon stars has sometimes been described as peculiar single stars resulting from past stellar mergers. Exotic binary interactions have been proposed to explain both the J-type stars (e.g., past merger event following a nova re-accretion event; \citealt{2013A&A...559A..66S}) and their suspected warmer progenitors, the early-R stars \citep{2007A&A...470..661I}. If such a merger -- between a white dwarf and a (post-)main-sequence star --  had taken place in the past evolution of the system, then it would probably have destroyed any substellar companion orbiting around the system. 
Hence, the two scenarios (J-type stars as post-merger products and long secondary periods as binary systems with a close low-mass companion) seem mutually incompatible, unless the putative substellar companion has formed in the aftermath of the merger or the objects were originally triple.

If instead the J-type spectral type is an imprint of a peculiar stellar evolutionary sequence (e.g., proton ingestion during the early-AGB phase or cold bottom burning), then the detection of a LSP provides constraints on their current evolutionary status. 
Indeed, the detection of standard LSP among J-type carbon stars implies at least that the J-type evolutionary path should lead to similar variability characteristics as standard semi-regular AGB stars experiencing the LSP phenomenon. J-Type carbon stars should thus already be thermally-pulsing AGB stars that have transitioned from sequence B to C', corresponding to the onset of dust production, self-excited pulsation and strong mass-loss rates \citep[$\sim 10^{-8} \, \rm M_\odot\, yr^{-1}$;][]  {LSP_mass_loss_2019MNRAS.484.4678M}.

\section{Conclusion} \label{sect:conclu}
In this article, we have reported the first high-resolution spectral monitoring of two J-type AGB stars: VZ Per and UV Cam. A LSP was discovered in their RV curve and is associated with a variation in the Balmer absorption profiles. Previous studies on oxygen-rich LSP stars have reported similar behavior, implying that these spectroscopic signatures of LSP are independent of the molecular content of the AGB photosphere. For one of the two stars, VZ Per, complementary photometric time series have confirmed the LSP. The dense spectral monitoring, spanning several LSP cycles, has allowed us to characterize the phase-dependent variability of the gas and dust density along the line of sight. 
In the binary framework, the prevailing scenario to explain the LSP, the periodic increase of gas and dust column density can be naturally explained by, and further support, the presence of a trailing structure -- such as a wake -- created behind a companion, and passing in front of the line of sight after the superior conjunction.

In conclusion, this study provided, for the first time, physical constraints on the circum-stellar/-companion environment of a star experiencing the LSP phenomenon. 
Detecting orbital motion in the RV curve of AGB stars is rare as it requires a long and dense time-series. Up to now, it was only restricted to the V~Hya system \citep{2024A&A...682A.143P}. This study enables the detection of two additional spectroscopic orbits among AGB stars, thereby increasing sensitivity to lower-mass companions.
The derived quantities and inferred orbital configurations aim to support the modeling effort by providing a direct comparison for hydrodynamic simulations of AGB stars with a nearby substellar companion. Complementary observations, especially those that probe the distribution of matter in the plane of the sky, such as high-resolution imaging, appear crucial for obtaining a complete view of these systems, including both their LSP nature and their peculiar J-type chemistry.

\begin{acknowledgements}

We thank the anonymous referee for their useful comments that
improved the quality of the paper. 
L.P. acknowledges support from the Olle Engkvist Foundation under project 229-0368. A.J. acknowledges support from the {\it Fonds de la Recherche Fondamentale Collective} (FNRS, F.R.F.C.) of Belgium
through grant PDR T.0115.23, and from FNRS-F.R.S. through grant CDR J.0100.21 ('ASTRO-HERMES').
Based on observations obtained with the HERMES spectrograph, which is supported by the Research Foundation - Flanders (FWO), Belgium, the Research Council of KU Leuven, Belgium, the Fonds National de la Recherche Scientifique (F.R.S.- FNRS), Belgium, the Royal Observatory of Belgium, the Observatoire de Genève, Switzerland, and the Thüringer Landessternwarte Tautenburg, Germany. We acknowledge all HERMES consortium observers who contributed to obtaining the time series. 
This work presents results from the European Space Agency (ESA) space mission Gaia. Gaia data are processed by the Gaia Data Processing and Analysis Consortium (DPAC). Funding for the DPAC is provided by national institutions, in particular the institutions participating in the Gaia MultiLateral Agreement (MLA). The Gaia mission website is \url{https://www.cosmos.esa.int/gaia}. The Gaia archive website is \url{https://archives.esac.esa.int/gaia}. 

Use was made of the Simbad database, operated at the CDS, Strasbourg, France, and of NASA's Astrophysics Data System Bibliographic Services. This research made use of Numpy, Matplotlib, SciPy, and Astropy, a community-developed core Python package for Astronomy \citep{Astropy_2018AJ....156..123A}. 
\end{acknowledgements}

\begingroup
\small
\bibliographystyle{aa}
\bibliography{references}
\endgroup


\appendix

\section{Log of observations}
Tables \ref{tab:RV_log} and \ref{tab:RV_log_UVCam} list the radial velocities from the HERMES spectrograph used in Sect. \ref{sect:time-series}.
\begin{table}[ht]
    \centering
    \small
    \begin{tabular}{rr|rr}
    \hline
\hline
    JD & RV & JD & RV  \\
    2400000+ & [km/s]  & 2400000+  &  [km/s]   \\
    \hline
55046.67 & -17.61   &56898.71 & -18.93  \\ 
55046.68 & -17.60   &56899.69 & -18.85  \\ 
55046.70 & -17.59   &56931.60 & -18.93  \\ 
55065.69 & -19.83    &56940.61 & -18.73 \\ 
55065.71 & -19.84   &56943.63 & -18.66  \\ 
55065.73 & -19.83    &56965.58 & -19.26 \\ 
55086.66 & -19.86    &56996.48 & -18.94 \\ 
55086.67 & -19.83    &57002.46 & -18.88 \\ 
55100.64 & -18.88    &57056.45 & -19.93 \\ 
55100.65 & -18.85    &57080.38 & -17.92 \\ 
55160.40 & -20.76    &57083.35 & -17.87 \\ 
55160.41 & -20.76    &57221.61 & -21.76 \\ 
55199.37 & -19.01    &57230.67 & -21.67 \\ 
55199.38 & -19.02    &57249.71 & -20.33 \\ 
55215.34 & -18.35    &57258.74 & -19.96 \\ 
55215.35 & -18.36    &57267.72 & -19.87 \\ 
55219.37 & -18.20    &57277.68 & -20.28 \\ 
55432.70 & -22.69    &57309.58 & -21.26 \\ 
55497.51 & -21.18    &57360.45 & -20.26 \\ 
55842.62 & -17.76    &57361.50 & -20.23 \\ 
55937.43 & -18.37    &57393.42 & -23.30 \\ 
55953.40 & -18.89    &57399.41 & -23.20\\ 
55959.41 & -20.27    &57624.64 & -22.30 \\ 
56017.36 & -19.28    &57642.65 & -22.52 \\ 
56108.72 & -18.08    &57675.62 & -23.78 \\ 
56139.65 & -21.45    &57716.57 & -22.61 \\ 
56146.68 & -21.93    &57737.35 & -21.66 \\ 
56189.63 & -19.08    &57777.33 & -20.66 \\ 
56247.52 & -21.35    &57787.35 & -20.99 \\ 
56307.41 & -21.53    &57977.73 & -20.38 \\ 
56322.35 & -22.29    &57985.71 & -19.84 \\ 
56482.71 & -22.00    &58004.73 & -18.32 \\ 
56483.67 & -22.07    &58058.57 & -20.43\\ 
56484.71 & -22.14    &58090.59 & -19.02 \\ 
56507.70 & -22.68    &58189.37 & -19.64 \\ 
56516.68 & -22.66    &58334.71 & -21.67 \\ 
56535.74 & -22.56    &58356.65 & -20.97 \\ 
56556.60 & -22.18    &58369.68 & -21.07 \\ 
56566.56 & -22.85    &58394.68 & -22.38 \\ 
56566.71 & -22.87    &58452.64 & -21.32 \\ 
56598.52 & -23.71    &58465.32 & -21.79 \\ 
56601.55 & -23.50    &58489.44 & -23.65 \\ 
56602.52 & -23.38    &58502.36 & -23.34 \\ 
56631.59 & -22.48    &58531.40 & -20.57\\ 
56662.47 & -23.17    &58558.40 & -21.01\\ 
56710.44 & -21.83    &58729.64 & -18.44 \\ 
56873.68 & -20.93    &58734.76 & -18.48 \\ 
56879.73 & -20.59    &58780.72 & -19.15 \\ 
    \end{tabular}
    \caption{log of observations with HERMES for VZ Per. }
    \label{tab:RV_log}
\end{table}

\begin{table}[ht]

    \centering
    \small
    \begin{tabular}{rr|rr}
    \hline
\hline
    JD & RV &  JD & RV  \\
    2400000+ & [km/s]  & 2400000+  &  [km/s]   \\
    \hline
55080.73 & -13.28 &56943.70 & -13.36  \\ 
55101.65 & -12.78  &56961.78 & -13.91  \\ 
55101.66 & -12.77  &57007.43 & -14.08  \\ 
55125.66 & -12.59  &57057.48 & -14.62  \\ 
55166.52 & -12.91  &57057.49 & -14.62 \\ 
55167.74 & -12.97 &57221.73 & -15.59  \\ 
55200.43 & -12.49 &57236.73 & -14.75 \\ 
55216.50 & -13.10 &57251.75 & -14.24 \\ 
55434.72 & -12.17 &57296.74 & -15.14  \\ 
55473.72 & -12.90 &57321.48 & -14.21 \\ 
55497.67 & -13.11  &57336.65 & -14.99 \\ 
55936.48 & -14.34 &57393.43 & -14.32  \\ 
55939.47 & -14.38 &57450.46 & -14.13 \\ 
55952.46 & -14.49 &57675.56 & -12.85 \\ 
55959.42 & -14.53 &57675.68 & -12.88 \\ 
56016.37 & -14.12 &57701.61 & -12.56 \\ 
56035.36 & -13.74 &57762.40 & -12.65  \\ 
56131.72 & -13.44 &57810.38 & -12.77  \\ 
56140.67 & -13.98 &57981.73 & -13.83  \\ 
56189.64 & -13.71 &58091.62 & -13.31 \\ 
56245.47 & -13.52 &58133.41 & -14.02 \\ 
56307.44 & -14.50 &58191.46 & -14.18 \\ 
56322.39 & -14.00 &58334.71 & -14.92 \\ 
56349.36 & -13.37 &58456.57 & -13.85 \\ 
56351.34 & -13.71 &58504.51 & -14.45 \\ 
56380.39 & -13.83 &58528.40 & -13.89 \\ 
56391.37 & -13.47 &58528.41 & -13.88 \\ 
56483.70 & -13.17 &58535.36 & -13.51 \\ 
56506.73 & -12.80 &58557.50 & -13.17 \\ 
56518.75 & -13.95 &58574.38 & -13.50\\ 
56538.74 & -13.47 &58603.36 & -12.35 \\ 
56555.69 & -12.79 &58773.72 & -12.40 \\ 
56566.73 & -12.31 &58781.64 & -12.71  \\ 
56598.58 & -13.13 &58857.54 & -12.08 \\ 
56630.69 & -12.60 &58885.53 & -11.65 \\ 
56630.69 & -12.59 &58905.45 & -11.41 \\ 
56701.44 & -12.35 &58922.42 & -11.70  \\ 
56751.37 & -13.34 &58922.43 & -11.66 \\ 
56874.74 & -13.03 &59095.74 & -12.71 \\ 
56879.74 & -13.44 &59095.75 & -12.71 \\ 
56931.61 & -13.14 &59126.60 & -12.09  \\ 
    \end{tabular}
    \caption{Same as Table \ref{tab:RV_log} for UV Cam.}
    \label{tab:RV_log_UVCam}
\end{table}

\section{Periodograms}

\begin{figure}[ht]
    \centering
    \includegraphics[width=\linewidth]{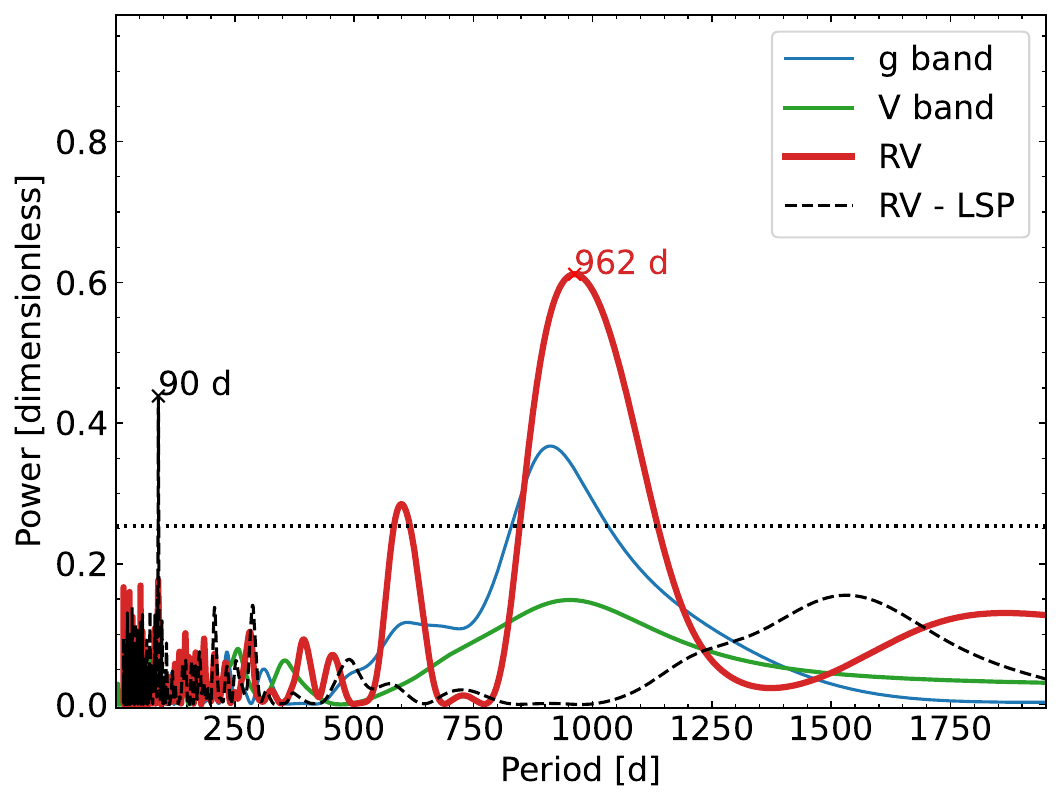}
    \includegraphics[width=\linewidth]{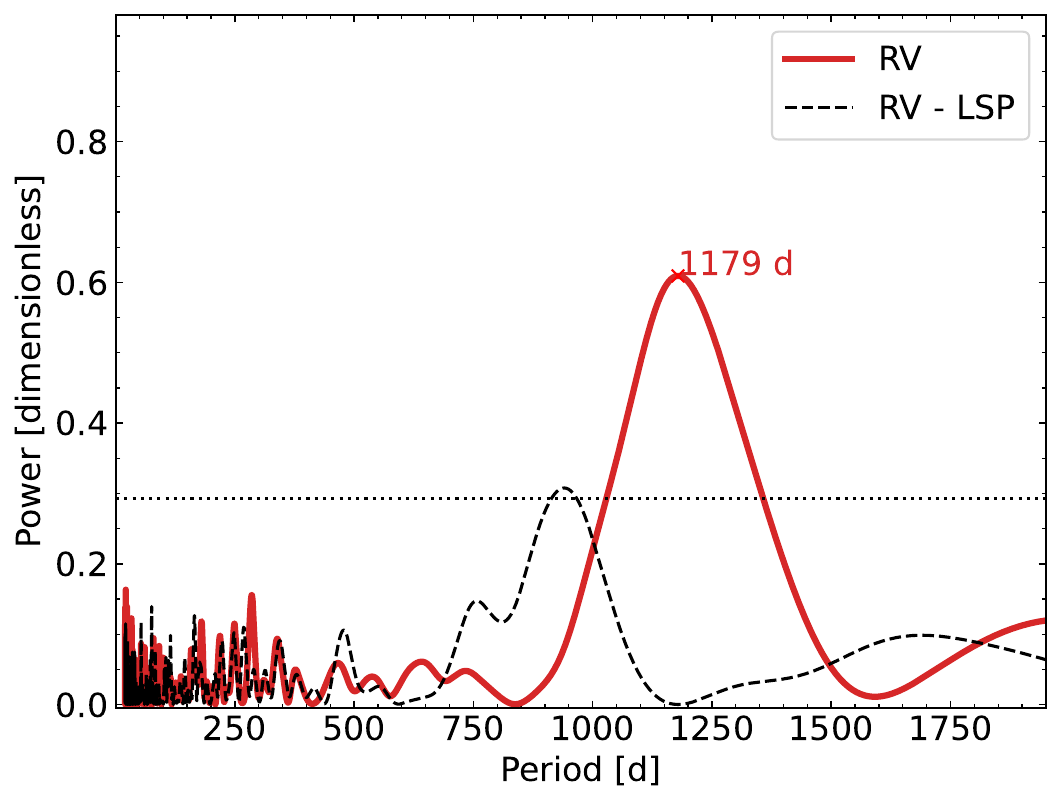}
    \caption{Periodograms. Top: VZ Per. Bottom: UV Cam. 'RV - LSP' corresponds to the RV-signal minus the best fit of a sinusoidal at the period of the maximum power.}
    \label{fig:Lombscargle}
\end{figure}

Figure \ref{fig:Lombscargle} displays the Lomb-Scargle periodograms \citep{LombScargle1982ApJ...263..835S} of the time series datasets displayed in Figs. \ref{fig:LC_VR_VZPer} and \ref{fig:LC_VR_UVCam} for VZ Per and UV Cam, respectively. 

\section{Spectral line variation}
\label{sect:spectral_line_overimposed}
\begin{figure}[th]
    \centering
    \includegraphics[width=0.49\linewidth]{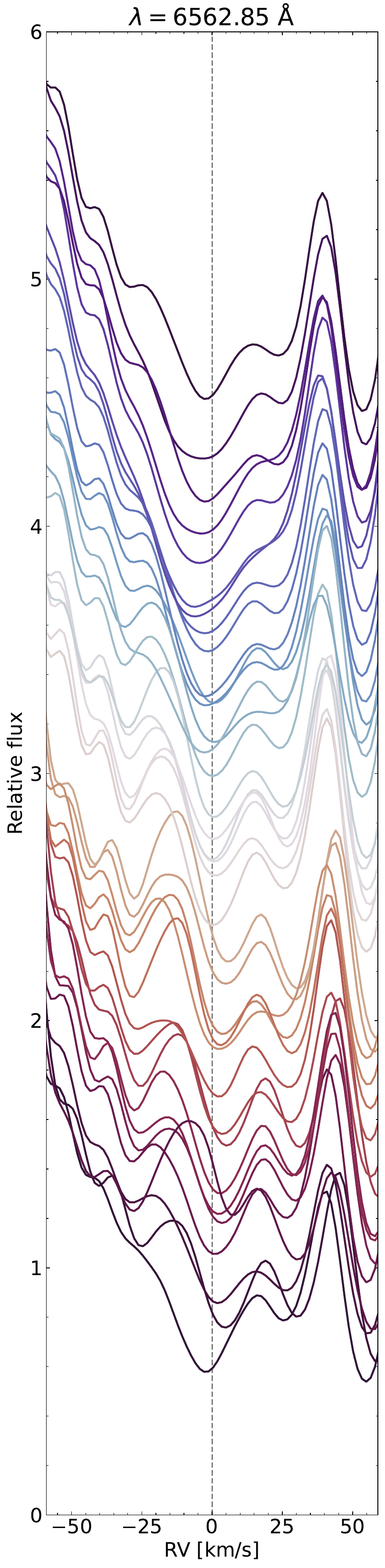}
    \includegraphics[width=0.495\linewidth]{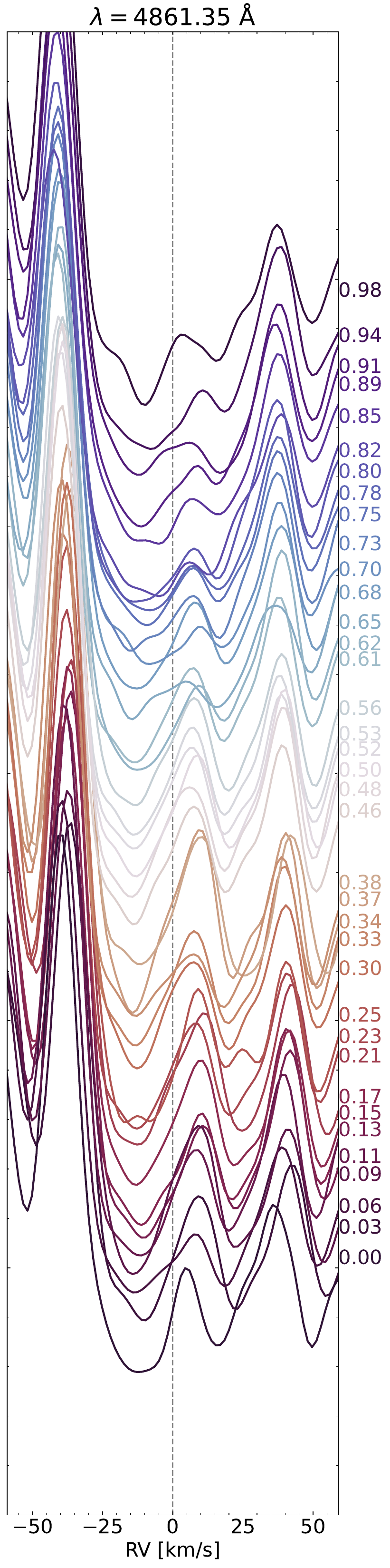}
    \caption{H$\alpha$ (leftmost panel) and H$\beta$ lines as a function of the orbital phase. The corresponding phases, $\phi$, are displayed on the rightmost axis. The flux values represent pseudo-continuum-normalized fluxes, taken as the median of the spectral window and incremented by steps of $4\times \Delta \phi$.}
    \label{fig:spectra_overimposedVZPer}
\end{figure}

\begin{figure}[th]
    \centering
    \includegraphics[width=0.49\linewidth]{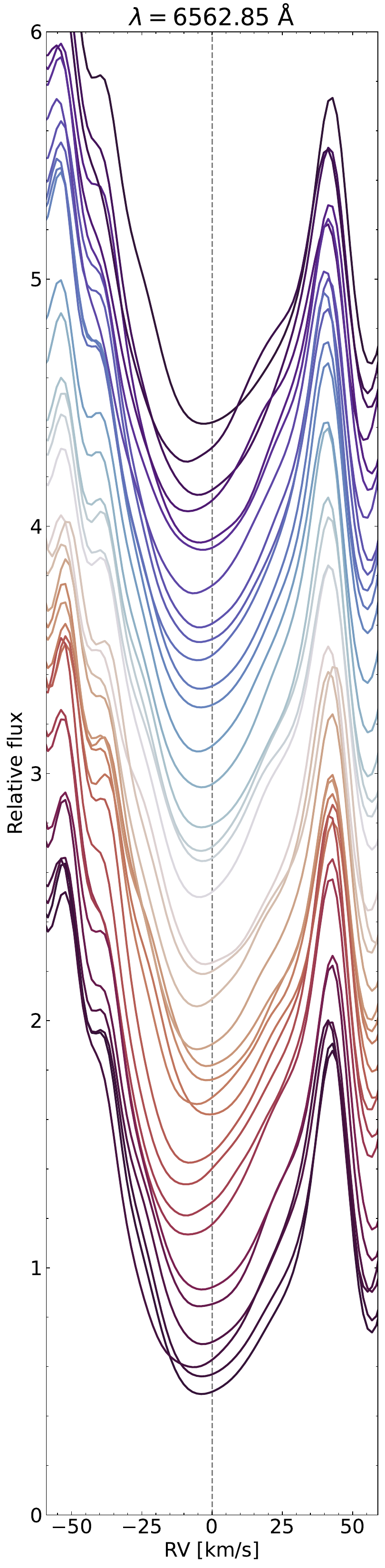}
    \includegraphics[width=0.495\linewidth]{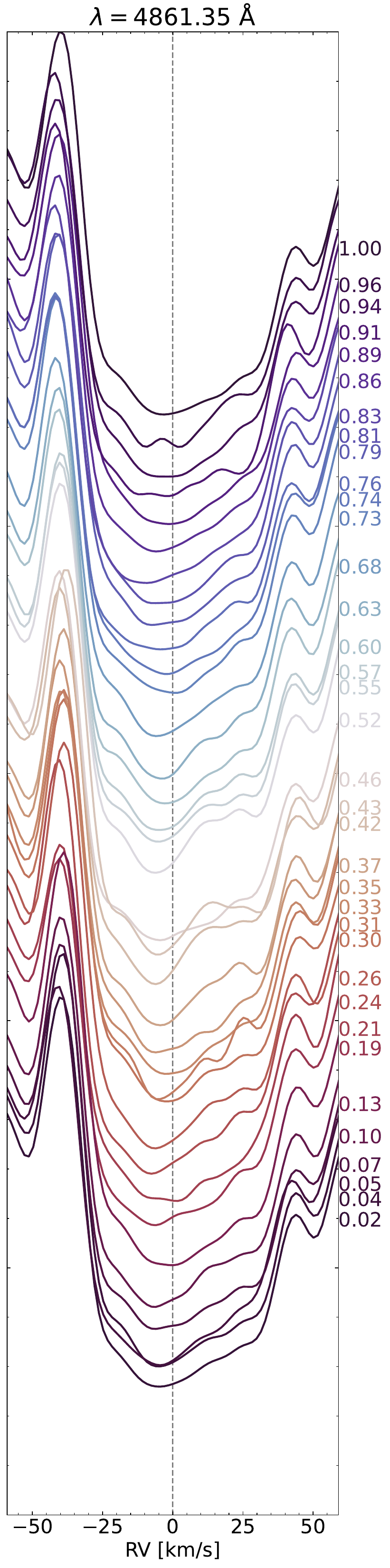}
    \caption{Same as Fig. \ref{fig:spectra_overimposedVZPer} but for UV Cam}
    \label{fig:spectra_overimposedUVCam}
\end{figure}

Spectral line variations are shown in Figs. \ref{fig:spectra_overimposedVZPer} and \ref{fig:spectra_overimposedUVCam} for VZ~Per and UV~Cam respectively. For the sake of readability, only spectra spaced by at least $\Delta \phi = 0.02$ are displayed. 

\section{Spectra of J-type carbon stars}
\label{sect:others_J-type_stars}
\begin{figure}[ht]
    \centering
    \includegraphics[width=\linewidth]{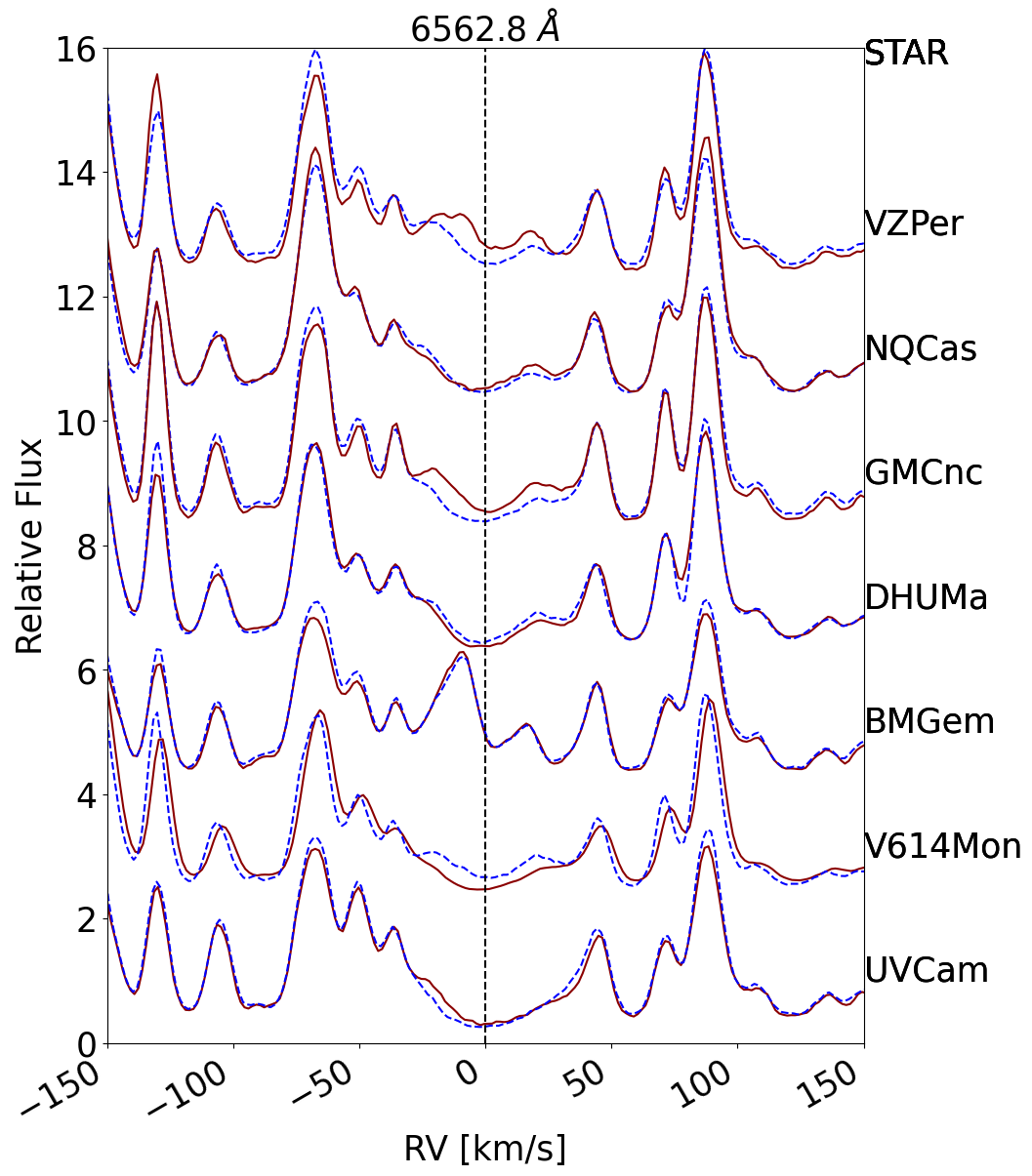}
    \caption{Comparison of H$\alpha$ spectral profiles for J-type stars obtained with the HERMES spectrograph. For each star, two profiles taken at different epochs are displayed (red continuous line and blue dashed line).}
    \label{fig:J_Halpha}
\end{figure}

\begin{figure}[ht]
    \centering
    \includegraphics[width=\linewidth]{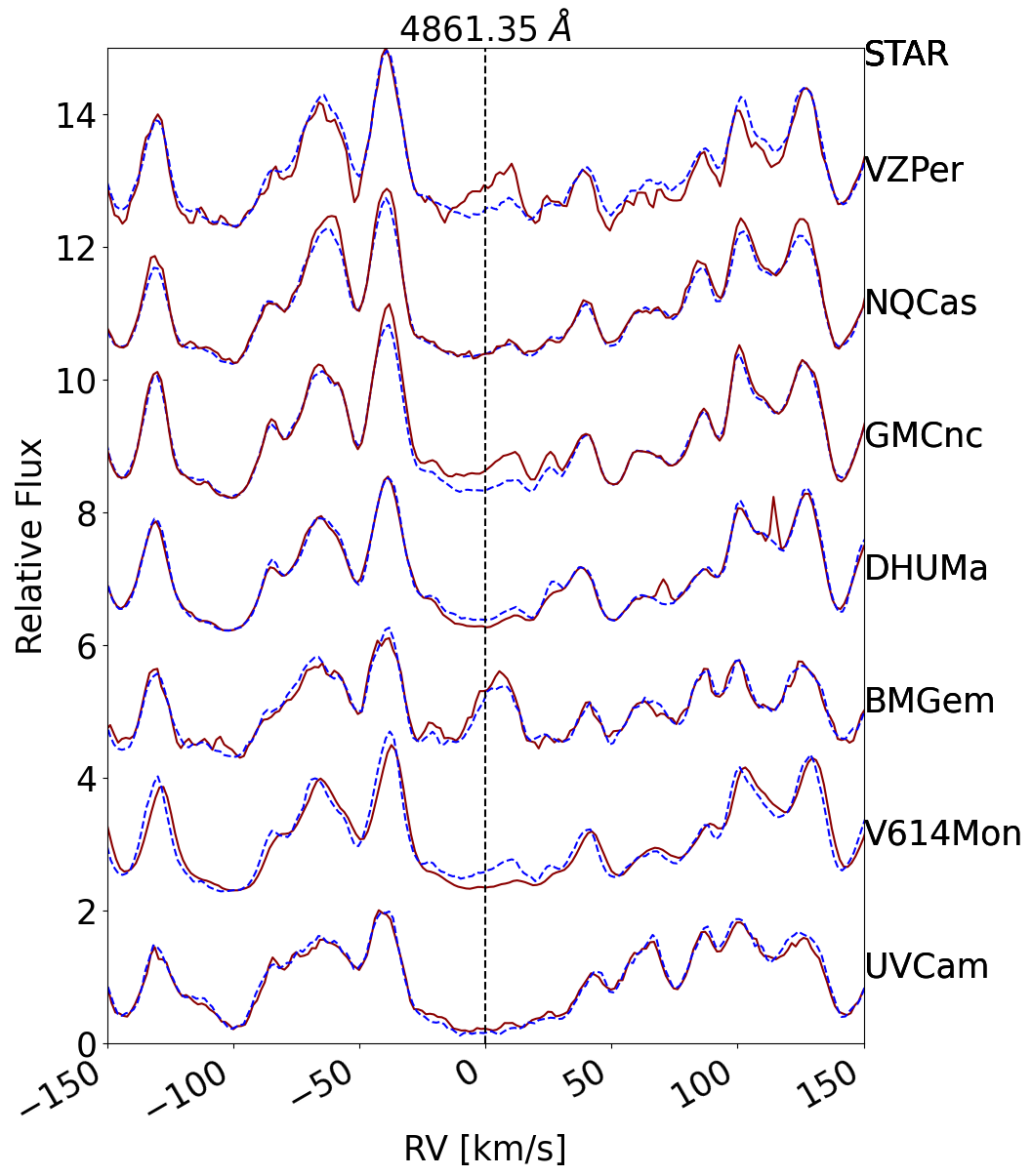}
    \caption{Same as Fig.\ref{fig:J_Halpha} but for H$\beta$.}
    \label{fig:J_Hbeta}
\end{figure}

Spectral line profiles of the first two Balmer transitions for a sample of J-type stars are plotted in Figs.~\ref{fig:J_Halpha} and \ref{fig:J_Hbeta}. Remarkably, the stars exhibit nearly identical profiles on the side of the Balmer line. The Balmer lines are observed in absorption, with some depth variations 
between the H$\alpha$ and H$\beta$ profiles, except for BM~Gem. 


\end{document}